\documentclass[aps,prb,superscriptaddress,footinbib,onecolumn,preprint]{revtex4} 
\usepackage{graphicx}
\usepackage{dsfont}
\usepackage[latin1]{inputenc}
\usepackage{version} 
\usepackage{color}
\usepackage{amsmath}
\usepackage{braket}
\usepackage{bm}
\usepackage{upgreek}
\usepackage{colortbl}
\usepackage{graphicx}
\usepackage{dcolumn}
\usepackage{bm}
\usepackage{bigints}
\usepackage{xcolor}  
\usepackage[pagebackref=false]{hyperref} 
\hypersetup{
    bookmarks=true,         
    unicode=false,          
    pdftoolbar=true,        
    pdfmenubar=true,        
    pdffitwindow=false,     
    pdfstartview={FitH},    
    pdftitle={A Coherent Spin-Photon Interface in Silicon},    
    pdfauthor={X. Mi},     
    pdfkeywords={keyword1} {key2} {key3}, 
    pdfnewwindow=true,      
    colorlinks=true,       
    linkcolor=red,          
    citecolor=blue,        
    filecolor=magenta,      
    urlcolor=cyan,           
}

\makeindex
\begin{document}
\title{A Coherent Spin-Photon Interface in Silicon} 
\author{X. Mi}
\affiliation{Department of Physics, Princeton University, Princeton, New Jersey 08544, USA}
\author{M. Benito}
\affiliation{Department of Physics, University of Konstanz, D-78464 Konstanz, Germany}
\author{S. Putz}
\affiliation{Department of Physics, Princeton University, Princeton, New Jersey 08544, USA}
\author{D. M. Zajac}
\affiliation{Department of Physics, Princeton University, Princeton, New Jersey 08544, USA}
\author{J.~M.~Taylor}
\affiliation{Joint Quantum Institute/NIST, College Park, Maryland 20742, USA}
\author{Guido~Burkard}
\affiliation{Department of Physics, University of Konstanz, D-78464 Konstanz, Germany}
\author{J.~R.~Petta}
\email{petta@princeton.edu}
\affiliation{Department of Physics, Princeton University, Princeton, New Jersey 08544, USA}
\date{\today}

\begin{abstract}
\textbf{Electron spins in silicon quantum dots are attractive systems for quantum computing due to their long coherence times and the promise of rapid scaling using semiconductor fabrication techniques. While nearest neighbor exchange coupling of two spins has been demonstrated, the interaction of spins via microwave frequency photons could enable long distance spin-spin coupling and ``all-to-all" qubit connectivity. Here we demonstrate strong-coupling between a single spin in silicon and a microwave frequency photon with spin-photon coupling rates $g_\text{s}/2 \pi$ $>$ 10 MHz. The mechanism enabling coherent spin-photon interactions is based on spin-charge hybridization in the presence of a magnetic field gradient. In addition to spin-photon coupling, we demonstrate coherent control of a single spin in the device and quantum non-demolition spin state readout using cavity photons. These results open a direct path toward entangling single spins using microwave frequency photons.}
\end{abstract}

\maketitle

Solid-state electron spins and nuclear spins are quantum mechanical systems that can be highly isolated from environmental noise, a virtue that endows them with coherence times as long as hours and establishes solid-state spins as one of the most promising quantum bits (qubits) for constructing a quantum processor \cite{Lyon_NatMat_2012,Saeedi2013}. On the other hand, this degree of isolation comes at a cost, as it poses difficulties for the spin-spin interactions needed to implement two-qubit gates. To date, most approaches have focused on achieving spin-spin coupling through the exchange interaction or the much weaker dipole-dipole interaction \cite{Petta_Science,Neumann1326,Dehollain2016}. Among existing classes of spin qubits, electron spins in gate-defined Si quantum dots (QDs) have the advantages of scalability due to mature fabrication technologies and low dephasing rates due to isotopic purification \cite{Petta_RevMod,RevModPhys.85.961}. Currently, Si QDs are capable of supporting fault-tolerant control fidelities for single-qubit gates and high fidelity two-qubit gates based on exchange \cite{VeldhorstM.2014,Takeda2016,Veldhorst2015,Zajac_Arxiv_2017,Watson_Arxiv_2017}. The recent demonstration of strong-coupling between the charge state of a Si electron and a single photon has also raised the prospect of spin-photon strong coupling, which could enable long-distance spin entanglement \cite{Mi_Science_2016}. Spin-photon coupling may be achieved by coherently hybridizing spin qubits with photons trapped inside microwave cavities, similar to cavity quantum electrodynamics (CQED) with atomic systems and circuit QED (cQED) with solid-state qubits \cite{PhysRevLett.68.1132,PhysRevLett.76.1800,Strong_Coupling_CooperPair,PhysRevA.69.042302,Mi_Science_2016,PhysRevX.7.011030}. Such an approach, however, is extremely challenging: the small magnetic moment of a single spin leads to magnetic-dipole coupling rates ranging from 10~--~150 Hz, which are far too slow compared with electron spin dephasing rates to enable a coherent spin-photon interface \cite{PhysRevA.69.042302,PhysRevLett.102.083602,PhysRevLett.105.140501,PhysRevLett.107.060502,Bienfait_Nature_2016,PhysRevLett.118.037701}.

In this Article, we resolve this outstanding challenge by using spin-charge hybridization to couple the electric field of a single photon to a single spin in Si \cite{PhysRevA.69.042302,Trif_PRB_2008,Hu_PRB_2012,Beaudoin_Nanotech_2017}. We measure spin-photon coupling rates $g_\text{s} / 2 \pi$ up to 11 MHz, nearly five orders of magnitude higher than typical magnetic-dipole coupling rates. These values of $g_\text{s} / 2 \pi$ exceed both the photon decay rate $\kappa / 2 \pi$ and the spin dephasing rate $\gamma_\text{s} / 2 \pi$, firmly anchoring our spin-photon system in the strong-coupling regime \cite{PhysRevLett.102.083602,Bienfait_Nature_2016,PhysRevLett.118.037701}.

Our coupling scheme consists of two stages of quantum state hybridization: First, a single electron is trapped within a gate-defined Si double quantum dot (DQD) having a large electric-dipole moment. A single photon confined within a microwave cavity hybridizes with the electron charge state through the electric-dipole interaction \cite{Walraff_2012_PRL,Petersson_Nature_2012}. Second, a micromagnet placed over the DQD hybridizes electron charge and spin by producing an inhomogeneous magnetic field \cite{Trif_PRB_2008,Hu_PRB_2012,Beaudoin_Nanotech_2017}. The combination of the electric-dipole interaction and spin-charge hybridization gives rise to a large effective spin-photon coupling rate. At the same time, the relatively low level of charge noise in the device ensures that the effective spin dephasing rate $\gamma_\text{s}$ remains below the coherent coupling rate $g_\text{s}$, a crucial criterion which has hampered a previous effort to achieve spin-photon strong coupling \cite{Viennot408}.

Beyond the demonstration of a coherent spin-photon interface, we also show that our device architecture is capable of single-spin control and readout. Single-spin rotations are electrically driven \cite{Kawakami.2014,Takeda2016} and the resulting spin state is detected through a dispersive phase shift in the cavity transmission, which reveals Rabi oscillations \cite{Petersson_Nature_2012}. Collectively, these results show that the spin-photon interface may serve as a fundamental building block for a Si-based spin quantum processor with the capacity for quantum non-demolition measurements and ``all-to-all'' connectivity \cite{Fowler_PRA_2012,IBM_NatComm_2015,Monroe_Nature_2016}. Moreover, the achievement of spin-photon strong coupling may allow highly coherent spin qubits to be entangled with other solid-state systems such as superconducting qubits \cite{Majer2007,Sillanpaa2007}.

\section*{Spin-Photon Interface}

The device enabling spin-photon strong coupling is shown in Fig.~\ref{fig:1}a and contains two gate-defined DQDs, which are fabricated using an overlapping aluminum gate stack (Fig.~\ref{fig:1}b). The gates are electrically biased to create a double-well potential that confines a single electron in the underlying natural-Si quantum well (Fig.~\ref{fig:1}c). A plunger gate (P2) on each DQD is connected to the center pin of a half-wavelength Nb superconducting cavity with a center frequency $f_\text{c} = 5.846$ GHz and quality factor $Q_\text{c}$ = 4,700 ($\kappa / 2\pi = f_\text{c}/Q_\text{c} = 1.3$ MHz), hybridizing the electron charge state with a single cavity photon through the electric-dipole interaction \cite{Walraff_2012_PRL,Mi_Science_2016,Mi_APL_2016}. Since the spin-photon coupling rate $g_\text{s}$ is directly proportional to the charge-photon coupling rate $g_\text{c}$ \cite{PhysRevA.69.042302,Burkard_PRB_2006,Trif_PRB_2008,Hu_PRB_2012,Jin_PRL_2012,Taylor_PRL_2013,Guido_RX_PRB_2015,Beaudoin_Nanotech_2017}, we have modified the cavity dimensions (inset of Fig.~\ref{fig:1}a) to achieve a high characteristic impedance $Z_\text{r}$ and therefore a high $g_\text{c}$ ($g_\text{c}$ $\propto \sqrt{Z_\text{r}}$) \cite{Mi_Science_2016,Mi_APL_2016}. To hybridize the trapped electron's charge state with its spin state, a Co micromagnet is fabricated near the DQD which generates an inhomogeneous magnetic field. For our device geometry, the magnetic field due to the Co micromagnet has a component along the $z$-axis, $B_\text{z}^\text{M}$, which is approximately constant for the DQD and a component along the $x$-axis which takes on a different average value of $B_\text{x,L}^\text{M}$ ($B_\text{x,R}^\text{M}$) for the left (right) dot, as illustrated by Fig.~\ref{fig:1}c. The relatively large field difference $B_\text{x,R}^\text{M} - B_\text{x,L}^\text{M} = 2 B_\text{x}^\text{M}$ leads to spin-charge hybridization which, when combined with charge-photon coupling, gives rise to spin-photon coupling\cite{Hu_PRB_2012,Beaudoin_Nanotech_2017}.

We first characterize the strength of the charge-photon interaction, since this sets the scale of the spin-photon interaction rate. For simplicity, only one DQD is active at a time for all of the measurements presented in this Article. The cavity is driven by a coherent microwave tone at frequency $f = f_\text{c}$ and power $P \approx -133$ dBm (corresponding to $\sim$1 intra-cavity photon). The normalized cavity transmission amplitude $A/A_\text{0}$ is displayed in Fig.~\ref{fig:1}d as a function of the voltages $V_\text{P1}$ and $V_\text{P2}$ on gates P1 and P2 of DQD 1, which reveals the location of the $(1,0) \leftrightarrow (0,1)$ interdot charge transition \cite{Mi_Science_2016}. Here $(N_1,N_2)$ denotes a charge state with the number of electrons in the left (P1) and right (P2) dot being $N_1$ and $N_2$, respectively. The charge-photon coupling rate is quantitatively estimated by measuring $A/A_0$ as a function of the DQD level detuning $\epsilon$ (Fig.~\ref{fig:1}e). By fitting the data to cavity input-output theory, we find $g_\text{c} / 2 \pi$ = 40 MHz and $2t_\text{c}/h = 4.9$ GHz, where $t_\text{c}$ is the interdot tunnel coupling and $h$ is Planck's constant \cite{Petersson_Nature_2012,Viennot408,Mi_Science_2016}. A charge dephasing rate $\gamma_\text{c} / 2 \pi = 35$ MHz is also estimated from the fit and independently confirmed using microwave spectroscopy \cite{Mi_Science_2016}. Fine control of the DQD tunnel coupling, which is critical for achieving spin-charge hybridization \cite{Hu_PRB_2012}, is shown in Fig.~\ref{fig:1}f where $2t_\text{c}/h$ is plotted as a function of the voltage $V_\text{B2}$ on the interdot barrier gate B2. A similar characterization of DQD 2 yields $g_\text{c} / 2 \pi$ = 37 MHz and $\gamma_\text{c} / 2 \pi = 45$ MHz at the $(1,0) \leftrightarrow (0,1)$ interdot charge transition. Due to the higher impedance of the resonator, the values of $g_\text{c}$ measured here are significantly larger than in previous Si DQD devices \cite{Mi_Science_2016,Mi_APL_2016}, which is helpful for achieving spin-photon strong coupling.

\section*{Single Spin-Photon Strong Coupling}

We now demonstrate strong-coupling between a single electron spin and a single photon, as evidenced by the observation of vacuum Rabi splitting. Vacuum Rabi splitting occurs when the transition frequency of a two-level atom $f_\text{a}$ is brought into resonance with a cavity photon of frequency $f_\text{c}$ \cite{PhysRevLett.68.1132,Strong_Coupling_CooperPair}. Light-matter hybridization leads to two ``vacuum-Rabi-split" peaks in the cavity transmission. For our single spin qubit, the transition frequency between two Zeeman-split spin states is $f_\text{a}$ $\approx$ $E_\text{Z} / h$, where the Zeeman energy $E_\text{Z}= g \mu_\text{B} B_\text{tot}$, and the approximate sign is due to spin-charge hybridization which slightly shifts the qubit frequency. Here $g$ is the electron $g$-factor, $\mu_\text{B}$ is the Bohr magneton and $B_\text{tot} = \sqrt{ ((B_\text{x,L}^\text{M}+B_\text{x,R}^\text{M})/2)^2+(B_\text{z}^\text{M}+B_\text{z}^\text{ext})^2 }$ is the total magnetic field. To bring $f_\text{a}$ into resonance with $f_\text{c}$, we vary the external magnetic field $B_\text{z}^\text{ext}$ along the $z$-axis while measuring the cavity transmission spectrum $A/A_0$ as a function of the drive frequency $f$, as shown in Fig.~\ref{fig:2}a. Vacuum Rabi splittings are clearly observed at $B_\text{z}^\text{ext} = -91.2$ mT and $B_\text{z}^\text{ext} = 92.2$ mT, indicating that $E_\text{Z} / h = f_\text{c}$ at these field values and the single spin is coherently hybridized with a single cavity photon. These measurements are performed on DQD 1, with $2t_\text{c}/h$ = 7.4 GHz and $\epsilon$ = 0. The dependences of $g_{\rm s}$ on $\epsilon$ and $t_\text{c}$ are thoroughly investigated below \cite{Hu_PRB_2012}. To compare these observations with theoretical expectations, $E_\text{Z} / h$ is shown in the lower panel of Fig.~\ref{fig:2}a as a function of $B_\text{tot}$, calculated using $g = 2$ for Si. The condition $E_\text{Z} / h = f_\text{c}$ occurs at $B_\text{tot} = 210$ mT, implying an intrinsic field of $\sim$120 mT is added by the micromagnet, comparable to values found in a previous experiment using a similar Co micromagnet design \cite{Takeda2016}.
 
To further verify the achievement of spin-photon strong coupling, we plot the cavity transmission spectrum at $B_\text{z}^\text{ext} = 92.2$ mT in Fig.~\ref{fig:2}b. The two normal mode peaks are separated by the vacuum Rabi frequency $2g_\text{s} / 2 \pi = 11.0$ MHz, giving an effective spin-photon coupling rate $g_\text{s} / 2 \pi = 5.5$ MHz. The photon decay rate at finite magnetic field is extracted by the linewidth of $A/A_0$ at $B_\text{z}^\text{ext} = 90.3$ mT where $E_\text{Z} / h \ll f_\text{c}$, yielding $\kappa / 2 \pi = 1.8$ MHz. A spin dephasing rate $\gamma_\text{s} / 2 \pi = 2.4$ MHz, with contributions from both charge noise and magnetic noise from the $^{29}$Si nuclei, is extracted from microwave spectroscopy in the dispersive regime with $2t_\text{c}/h$ = 7.4 GHz and $\epsilon$ = 0 (see Fig.~\ref{fig:4}b for data at a higher $t_\text{c}$), confirming that the strong-coupling regime $g_\text{s} > \gamma_\text{s}, \kappa$ has been reached. It is remarkable that the spin-photon coupling rate obtained here is more than four orders of magnitude larger than currently achievable rates using direct magnetic-dipole coupling to lumped element superconducting resonators \cite{PhysRevLett.118.037701,Bertet_Arxiv_2017}.

The local magnetic field generated using Co micromagnets is very reproducible, as evidenced by examining the other DQD in the cavity. Measurements on DQD 2 show vacuum Rabi splittings at $B_\text{z}^\text{ext}$ = $\pm 92.6$ mT (insets to Fig.~\ref{fig:2}a). The spin-photon coupling rate and spin dephasing rate are determined to be $g_\text{s} / 2 \pi = 5.3$ MHz and $\gamma_\text{s} / 2 \pi = 2.4$ MHz respectively (Fig.~\ref{fig:2}c). These results are highly consistent with DQD 1, which we focus on for the rest of this Article.

\section*{Electrical Control of Spin-Photon Coupling}

For quantum information applications, it is desirable to rapidly turn qubit-cavity coupling on for quantum state transfer, and off for qubit state preparation. Fast control of the coupling rate is often accomplished by quickly modifying the qubit-cavity detuning $f_\text{a} - f_\text{c}$. Practically, such tuning can be achieved by varying the qubit transition frequency $f_\text{a}$ with voltage or flux pulses \cite{Majer2007,Sillanpaa2007}, or by using a tunable cavity \cite{PhysRevX.7.011030}. These approaches are not directly applicable for control of the spin-photon coupling rate since $f_\text{a}$ primarily depends on magnetic fields that are difficult to vary on nanosecond timescales. In this section, we show that control of the spin-photon coupling rate may be achieved electrically by tuning $\epsilon$ and $t_\text{c}$ \cite{Jin_PRL_2012}. 

We first investigate the $\epsilon$ dependence of $g_{\rm s}$. Figure~\ref{fig:3}a shows measurements of $A/A_0$ as a function of $B_\text{z}^\text{ext}$ and $f$ for $\epsilon$ = 0, 20 $\mu$eV and 40 $\mu$eV. At $\epsilon$ = 20 $\mu$eV, vacuum Rabi splitting is observed at $B_\text{z}^\text{ext} = 92.1$ mT with a spin-photon coupling rate $g_\text{s} / 2 \pi = 1.0$ MHz that is significantly lower than the value of $g_\text{s} / 2 \pi$ = 5.5 MHz obtained at $\epsilon$ = 0. At $\epsilon$ = 40 $\mu$eV, only a small dispersive shift is observed in the cavity transmission spectrum at $B_\text{z}^\text{ext} = 91.8$ mT, suggesting further decrease in $g_\text{s}$. These observations are qualitatively understood by considering that at $\epsilon$ = 0 the electron is delocalized across the DQD and forms molecular bonding(anti-bonding) charge states $\ket{-} (\ket{+})$ (Fig.~\ref{fig:3}c). In this regime, the cavity electric field leads to a large displacement of the electron wavefunction. Consequently, the electron spin experiences a large oscillating magnetic field, resulting in a substantial spin-photon coupling rate. In contrast, with $|\epsilon| \gg t_\text{c}$, the electron is localized within one dot and it is natural to work with a basis of localized electronic wavefunctions $\ket{L} (\ket{R})$ where $L (R)$ corresponds to the electron being in the left(right) dot (Fig.~\ref{fig:3}c). In this effectively single-dot regime, the displacement of the electron wavefunction by the cavity electric field is estimated to be of order $\sim$ 1 nm, greatly suppressing the spin-photon coupling mechanism \cite{Nadj_Nature_2010}. More quantitatively, theory predicts $g_\text{s} \approx \frac{g \mu_\text{B} B_\text{x}^\text{M}}{E_\text{orb}} g_\text{c}$ for a single QD and $g_\text{s} \approx \frac{g \mu_\text{B} B_\text{x}^\text{M}}{t_\text{c}} g_\text{c}$ for a DQD at $\epsilon$ = 0, where $E_\text{orb}$ is the orbital energy of a single QD \cite{Trif_PRB_2008,Hu_PRB_2012,Beaudoin_Nanotech_2017}. With $E_\text{orb} = 2.5$ meV, we therefore expect a factor of $E_\text{orb} / t_\text{c} \approx 200$ improvement in the spin-photon coupling rate at $\epsilon$ = 0 compared to $|\epsilon| \gg t_\text{c}$ \cite{Dave_DQD_APL}. These measurements highlight the important role of charge hybridization in the DQD. 

Additional electric control of $g_{\rm s}$ is enabled by voltage-tuning $t_\text{c}$ (Fig.~\ref{fig:1}f). Figure~\ref{fig:3}b shows $g_\text{s} / 2 \pi$ and $\gamma_\text{s} / 2 \pi$ as a function of $2t_\text{c}/h$ at $\epsilon = 0$, as extracted from vacuum Rabi splitting measurements and microwave spectroscopy of the ESR transition linewidth (Fig.~\ref{fig:4}b). Both rates rapidly increase as $2t_\text{c}/h$ approaches the Larmor precession frequency $E_\text{Z} / h \approx 5.8$ GHz, and a spin-photon coupling rate as high as $g_\text{s} / 2 \pi = 11.0$ MHz is found at $2t_\text{c}/h = $ 5.2 GHz. These trends are consistent with the DQD energy level spectrum shown in Fig.~\ref{fig:3}c \cite{Hu_PRB_2012,Beaudoin_Nanotech_2017}. Here $\uparrow (\downarrow)$ denotes an electron spin that is aligned(anti-aligned) with $B_\text{z}^\text{ext}$. With $2t_\text{c} / h \gg E_\text{Z} / h$ and $\epsilon = 0$, the two lowest energy levels are $\ket{-, \downarrow}$ and $\ket{-, \uparrow}$ and the electric-dipole coupling to the cavity field is small. As $2t_\text{c}$ is reduced and made comparable to $E_{\rm Z}$, the ground state remains $\ket{-, \downarrow}$ but the excited state becomes an admixture of $\ket{-, \uparrow}$ and $\ket{+, \downarrow}$ due to the magnetic field gradient $B_\text{x,R}^\text{M} - B_\text{x,L}^\text{M} = 2 B_\text{x}^\text{M}$ and the small energy difference between the states. The quantum transition that is close to resonance with $E_{\rm Z}$ is now partially composed of a change in charge state from $-$ to $+$, which responds strongly to the cavity electric field and gives rise to larger values of $g_\text{s}$. For $2t_\text{c} / h < E_\text{Z} / h$, a decrease in $t_\text{c}$ increases the energy difference between $\ket{-, \uparrow}$ and $\ket{+, \downarrow}$ which reduces their hybridization and results in a smaller $g_\text{s}$. We note that hybridization with charge states increases the susceptibility of the spin to charge noise, and results in an effective spin dephasing rate $\gamma_\text{s}$ that is a strong function of $t_\text{c}$ as well (see Fig.~\ref{fig:3}b). Theoretical predictions of $g_\text{s}$ and $\gamma_\text{s}$ as a function of $2t_\text{c}/h$, based on measured values of $g_\text{c}$ and $\gamma_\text{c}$ (Fig.~\ref{fig:1}e), are in good agreement with the data (Fig.~\ref{fig:3}b). The discrepancy in the fit of $\gamma_\text{s}$ may be due to an incomplete theoretical treatment of the noise mechanisms in the device, which is beyond the scope of this work (see Methods). The electric control of spin-photon coupling demonstrated here allows the spin qubit to quickly switch between regimes with strong coupling to the cavity, and idle regimes where the spin-photon coupling rate and susceptibility to charge noise are small.

\section*{Quantum Control and Dispersive Readout of a Single Spin}

The preceding measurements demonstrate the ability to coherently couple a single electron spin to a single photon, potentially enabling long-range spin-spin couplings \cite{Majer2007,Sillanpaa2007}. For the device to serve as a building block of a quantum processor, it is also necessary to deterministically prepare, control, and read out the spin state of the trapped electron. We first induce spin transitions by driving gate P1 with a continuous microwave tone of frequency $f_\text{s}$ and power $P_\text{s} = -106$ dBm. When $f_\text{s} \approx E_\text{Z} / h$, the excited state population of the spin qubit $P_\uparrow$ increases and the ground state population $P_\downarrow$ decreases. In the dispersive regime, where the qubit-cavity detuning $\Delta / 2 \pi \approx E_\text{Z} / h - f_\text{c}$ satisfies $|\Delta / 2 \pi| \gg g_\text{s} / 2 \pi$, the cavity transmission experiences a phase response $\Delta \phi \approx \tan^{-1} (2g_\text{s}^2/\kappa \Delta)$ for a fully saturated ($P_\uparrow = 0.5$) qubit \cite{Schuster_ACStark_2005,Mi_Science_2016}. It is therefore possible to measure the spin state of a single electron by probing the cavity transmission. As a demonstration, we spectroscopically probe the electron spin resonance (ESR) transition by measuring $\Delta \phi$ as a function of $f_\text{s}$ and $B_\text{z}^\text{ext}$ in Fig.~\ref{fig:4}a. These data are acquired with $2t_\text{c}/h$ = 9.5 GHz and $\epsilon$ = 0. The ESR transition is clearly visible as a narrow feature with $\Delta \phi \neq 0$ that shifts to higher $f_{\rm s}$ with increasing $B_{\rm z}^{\rm ext}$. $\Delta \phi$ also changes sign as $B_{\rm z}^{\rm ext}$ increases, consistent with the sign change of the qubit-cavity detuning $\Delta$ when the Larmor precession frequency $E_\text{Z} / h$ exceeds $f_\text{c}$. The nonlinear response in the small region around $B_\text{z}^\text{ext} = 92$ mT is due to the breakdown of the dispersive condition $|\Delta / 2 \pi| \gg g_\text{s} / 2 \pi$.

Finally, we demonstrate coherent single spin control and dispersive spin state readout. For these measurements $2t_\text{c} / h =$ 11.1~GHz and $\epsilon = 0$, resulting in a spin-photon coupling rate $g_\text{s} / 2 \pi = 1.4$~MHz. The external field is fixed at $B_\text{z}^\text{ext}$ = 92.18~mT, ensuring the system is in the dispersive regime with $\Delta / 2\pi = 14$~MHz $\gg$ $g_\text{s} / 2 \pi$. A measurement of $\Delta \phi (f_\text{s})$ in the low power limit (Fig.~\ref{fig:4}b) yields a Lorentzian line shape with a full-width-at-half-maximum of 0.81~MHz, corresponding to a low spin dephasing rate $\gamma_\text{s} / 2 \pi = 0.41$ MHz \cite{Schuster_ACStark_2005,Mi_Science_2016}. Qubit control and measurement are achieved using the pulse sequence illustrated in Fig.~\ref{fig:4}c: Starting with a spin-down state $\ket{\downarrow}$ at $\epsilon = 0$, the DQD is pulsed to a large detuning $\epsilon' = 70$~$\mu$eV which decouples the spin from the cavity. A microwave burst with frequency $f_\text{s}$ = 5.874~GHz, power $P_\text{s} = -76$ dBm, and duration $\tau_\text{B}$ is subsequently applied to P1 to drive a spin rotation \cite{Petersson_Nature_2012,Kawakami.2014,Takeda2016}. The DQD is then adiabatically pulsed back to $\epsilon = 0$ for a fixed measurement time  $T_\text{M}$ for dispersive readout. To reinitialize the qubit, we choose $T_\text{M} = 20$ $\mu$s $\gg$ $T_1 (\epsilon = 0)$, where $T_1 (\epsilon = 0) = 3.2$ $\mu$s is the spin relaxation time measured at $\epsilon = 0$. Figure~\ref{fig:4}d displays the time-averaged $\Delta \phi$ as a function of $\tau_\text{B}$. We observe coherent single spin Rabi oscillations with a Rabi frequency $f_\text{R}$ = 6 MHz. In contrast with readout approaches that rely on spin-dependent tunneling, our dispersive cavity-based readout performs a quantum non-demolition measurement \cite{Elzerman_Readout,Kawakami.2014,Takeda2016}. In addition to enabling single spin-photon coupling, our device is capable of preparing, controlling, and reading out single spin states.

\section*{Conclusion}

In conclusion, we have realized a coherent spin-photon interface where a single spin in a Si DQD is strongly coupled to a microwave photon through the combined effects of the electric-dipole interaction and spin-charge hybridization. Spin-photon coupling rates up to 11 MHz are measured in the device, exceeding magnetic-dipole coupling rates by nearly five orders of magnitude. The spin dephasing rate is strongly dependent on 2$t_{\rm c}/h$ and ranges from 0.4 -- 6 MHz, limited by a combination of charge noise and remnant nuclear field fluctuations in natural-Si. All-electric control of spin-photon coupling and coherent manipulation of the spin state are demonstrated, along with quantum non-demolition readout of the single spin through its dispersive interaction with the microwave cavity. These results may enable the construction of an ultra-coherent spin quantum computer having photonic interconnects and readout channels, with capacity for surface codes, ``all-to-all'' connectivity, and easy integration with other solid-state quantum systems such as superconducting qubits \cite{Majer2007,Sillanpaa2007,Fowler_PRA_2012,IBM_NatComm_2015,Monroe_Nature_2016}.

\newpage

\begin{figure}[!ht]
\centering
\includegraphics[width=0.7\textwidth]{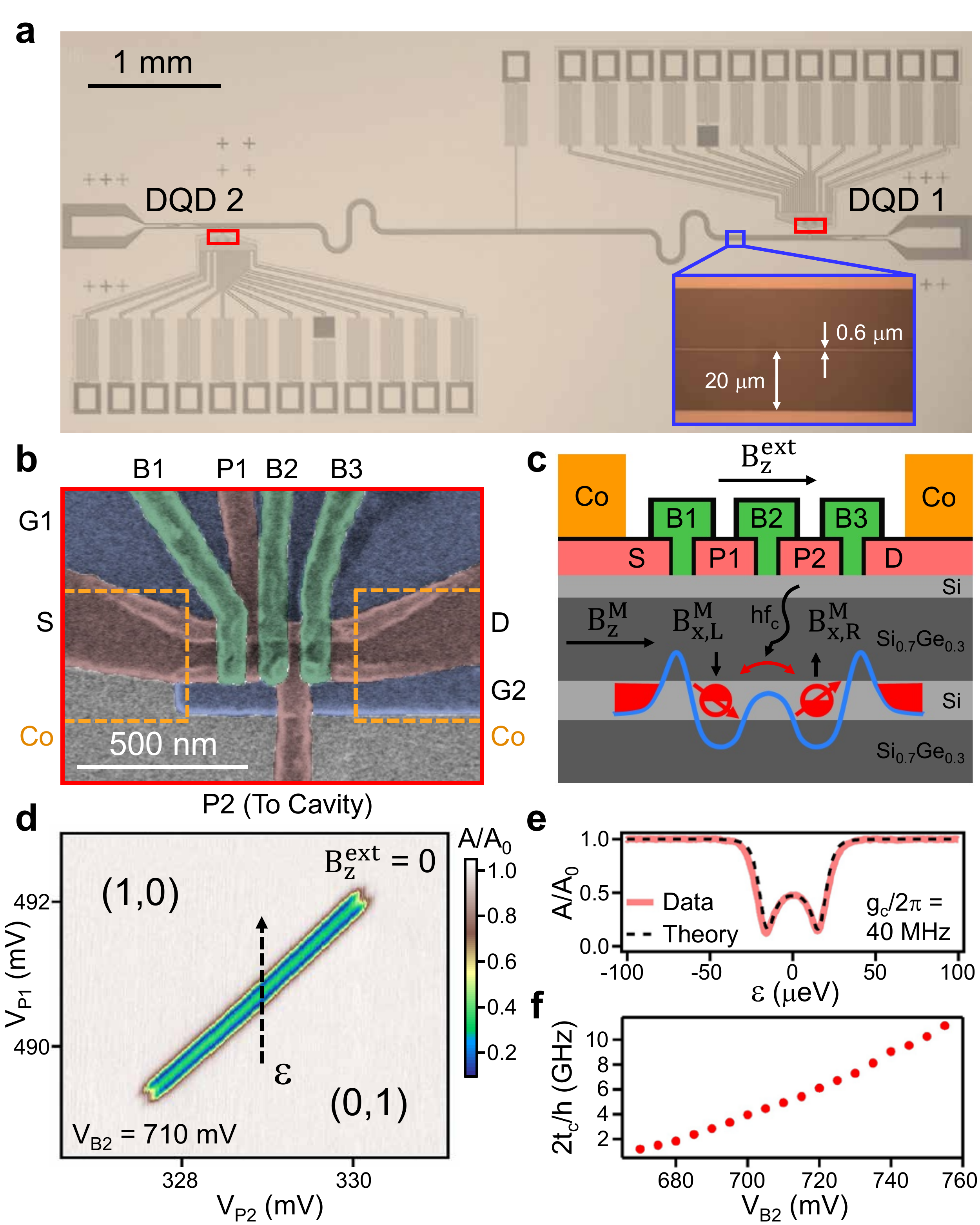}
\linespread{1.0}
\caption{\textbf{Spin-Photon Interface.} \textbf{a,} Optical image of the superconducting microwave cavity. Inset shows an optical image of the cavity center pin (0.6 $\mu$m) and vacuum gap (20 $\mu$m). \textbf{b,} Tilted-angle false-color SEM of a DQD. Gate electrodes are labeled as G1, G2, S, D, B1, P1, B2, P2 and B3. The Co micromagnet location is indicated by the orange dashed lines. \textbf{c,} Schematic cross-sectional view of the DQD device. \textbf{d,} Cavity transmission amplitude $A/A_0$ at $f = f_\text{c}$ near the $(1,0) \leftrightarrow (0,1)$ interdot transition for DQD 1. Dashed arrow denotes the DQD detuning parameter $\epsilon$. \textbf{e,} $A/A_0$ as a function of $\epsilon$ with $V_\text{B2}$ = 710 mV, and a fit to cavity input-output theory. \textbf{f,} $2t_\text{c}/h$ as a function of $V_\text{B2}$ for DQD 1, obtained by measuring $A (\epsilon) /A_0$ at different values of $V_\text{B2}$.}
\label{fig:1}
\end{figure}

\newpage

\begin{figure*}[!ht]
\centering
\includegraphics[width=0.8\textwidth]{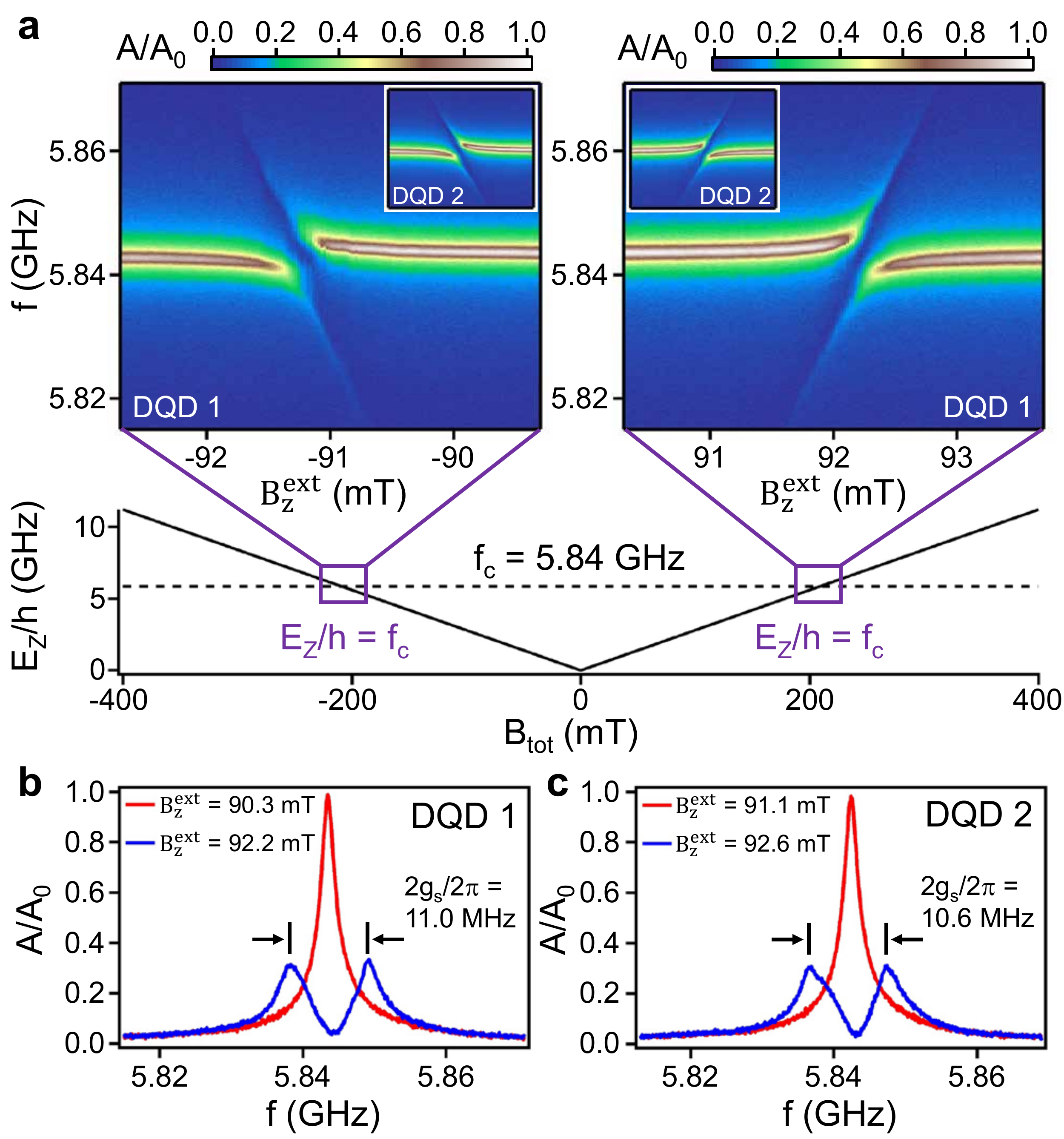}
\linespread{1.0}
\caption{\textbf{Single Spin-Photon Strong Coupling.} \textbf{a,} $A/A_0$ as a function of $f$ and $B_\text{z}^\text{ext}$ for DQD 1. Insets show data from DQD 2 at the same values of $t_\text{c}$ and $\epsilon$, and plotted over the same range of $f$. $B_{\rm z}^{\rm ext}$ ranges from -94 mT to -91.1 mT (91.1 mT to 94 mT) for the left (right) inset. Lower panel shows the Larmor precession frequency $E_\text{Z} / h$ as a function of $B_\text{tot}$, calculated for a single electron spin in Si. \textbf{b,} $A/A_0$ as a function of $f$ for DQD 1 at $B_\text{z}^\text{ext}$ = 90.3 mT and 92.2 mT. \textbf{c,} $A/A_0$ as a function of $f$ for DQD 2 at $B_\text{z}^\text{ext}$ = 91.1 mT and 92.6 mT.}
\label{fig:2}
\end{figure*} 

\newpage

\begin{figure*}[!ht]
\centering
\includegraphics[width=0.8\textwidth]{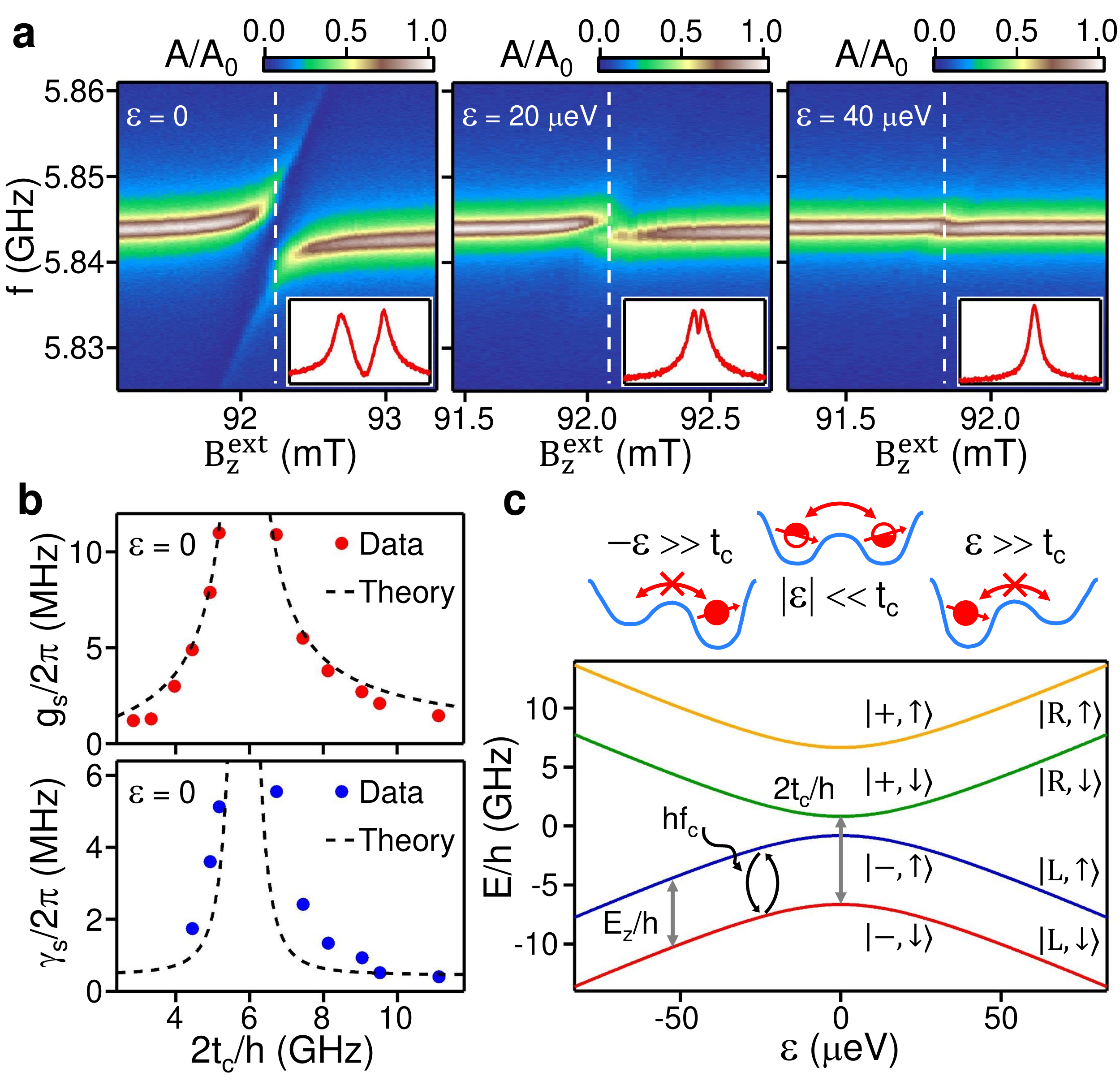}
\linespread{1.0}
\caption{\textbf{Electrical Control of Spin-Photon Coupling.} \textbf{a,} $A/A_0$ as a function of $f$ and $B_\text{z}^\text{ext}$ at $\epsilon$ = 0, $\epsilon$ = 20 $\mu$eV and $\epsilon$ = 40 $\mu$eV, with $2t_\text{c}/h$ = 7.4 GHz. Insets show $A/A_0$ as a function of $f$ at values of $B_\text{z}^\text{ext}$ indicated by the white dashed lines. \textbf{b,} Spin-photon coupling rate $g_\text{s} / 2 \pi$ and spin decoherence rate $\gamma_\text{s} / 2 \pi$ as a function of $2t_\text{c}/h$, with $\epsilon = 0$. The dashed lines show theory predictions. \textbf{c,} DQD energy levels as a function of $\epsilon$, calculated with $B_\text{z}^\text{ext} + B_\text{z}^\text{M}$ = 209 mT, $B_\text{x}^\text{M}$ = 15 mT and $2t_\text{c}/h$ = 7.4 GHz.  }
\label{fig:3}
\end{figure*} 

\newpage

\begin{figure}[!ht]
\centering
\includegraphics[width=0.8\textwidth]{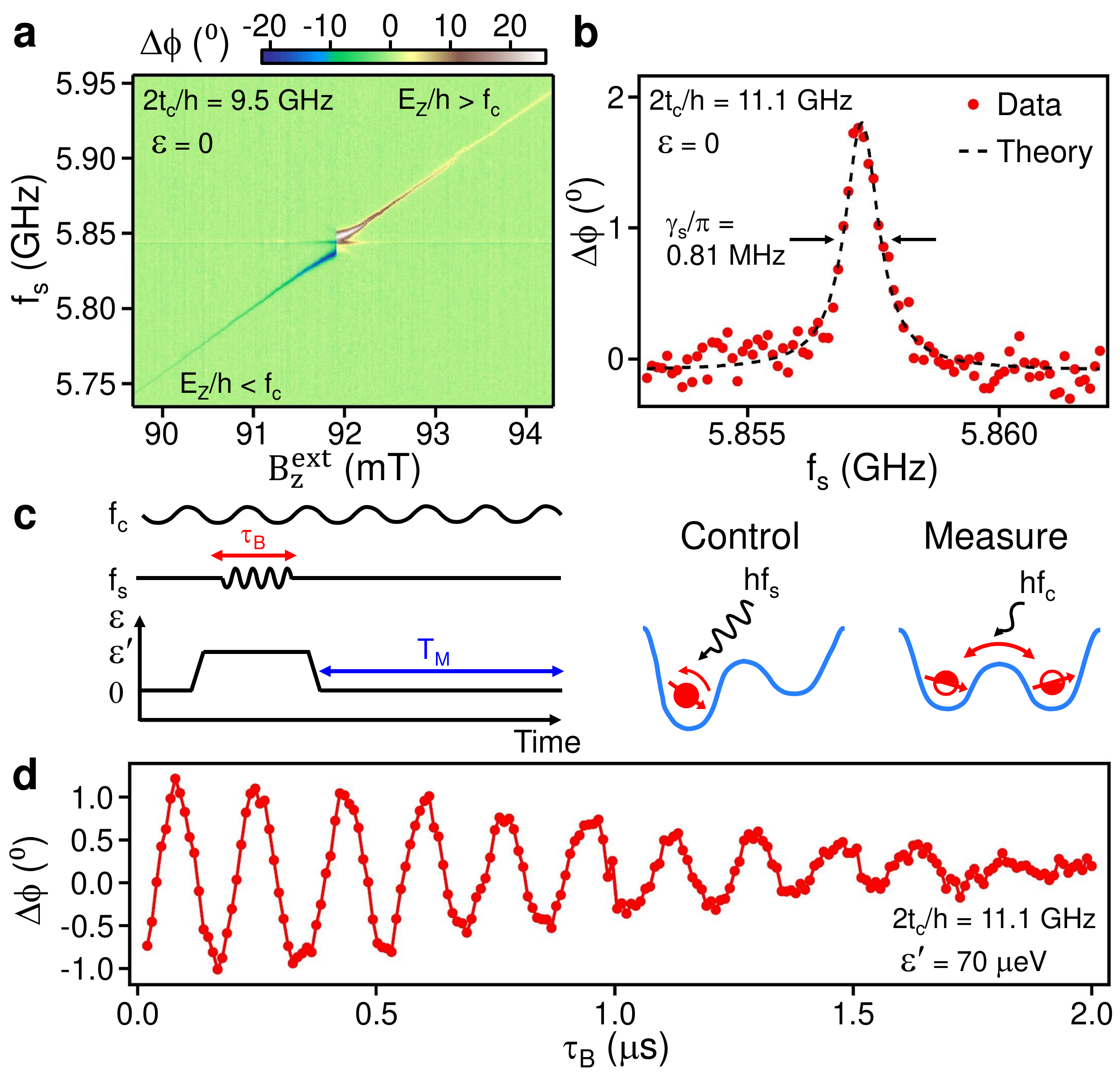}
\linespread{1.0}
\caption{\textbf{Quantum Control and Dispersive Readout of a Single Spin.} \textbf{a,} Cavity phase response $\Delta \phi$ at $f = f_\text{c}$ when gate P1 is continuously driven at a variable frequency $f_\text{s}$ and power $P_\text{s}$ = -106 dBm. A background phase response, obtained by measuring $\Delta \phi (B_\text{z}^\text{ext})$ in the absence of a microwave drive on P1, is subtracted from each column of the data to correct for slow drifts in the microwave phase. \textbf{b,} ESR line as measured in $\Delta \phi (f_\text{s})$ at $2t_\text{c}/h$ = 11.1 GHz, $\epsilon$ = 0, $B_\text{z}^\text{ext}$ = 92.18 mT and $P_\text{s}$ = -123 dBm. Dashed line shows a fit to a Lorentzian with FWHM $\gamma_\text{s} / \pi$ = 0.81 MHz. \textbf{c,} Schematic showing the experimental sequence for coherent spin control and measurement. \textbf{d,} $\Delta \phi$ as a function of $\tau_\text{B}$, showing single spin Rabi oscillations. Solid line is a guide to the eye.} 
\label{fig:4}
\end{figure}

\newpage
\clearpage

\textbf{~\\\large{Methods}}

\textbf{Device fabrication and measurement} The Si/SiGe heterostructure consists of a 4 nm thick Si cap, a 50 nm thick Si$_{0.7}$Ge$_{0.3}$ spacer layer, a 8 nm thick natural-Si quantum well, and a 225 nm thick Si$_{0.7}$Ge$_{0.3}$ layer on top of a linearly graded Si$_{1-x}$Ge$_{x}$ relaxed buffer substrate. Design and fabrication details for the superconducting cavity and DQDs are described elsewhere \cite{Mi_APL_2016}. The $\sim$200 nm thick Co micromagnet is defined using electron beam lithography and liftoff. In contrast with earlier devices, the gate filter for P1 has been changed to an $L_1$-$C$-$L_2$ filter, with $L_1 = 4$ nH, $C = 1$ pF and $L_2 = 12$ nH  \cite{Mi_APL_2016}. This three-segment filter allows microwave signals below $2.5$ GHz to pass with $<3$ dB of attenuation.

All data are acquired in a dilution refrigerator with a base temperature of 10 mK and electron temperature $T_\text{e} = 60$ mK. The measurements of cavity transmission amplitude and phase response in Fig.~\ref{fig:1} and Fig.~\ref{fig:4} are performed using a homodyne detection scheme similar to previous works \cite{Mi_Science_2016,Mi_APL_2016}. The measurements of cavity transmission spectra in Fig.~\ref{fig:2} and Fig.~\ref{fig:3} are performed using a network analyzer. The microwave drive applied to P1 in Fig.~\ref{fig:4} is provided by a vector microwave source and the detuning pulses are generated by an arbitrary waveform generator, which also controls the timing of the microwave burst in Fig.~\ref{fig:4}d.

To maximize the magnetization of the Co micromagnet and minimize hysteresis, data at positive/negative external applied magnetic fields in Fig.~\ref{fig:2}a are taken after $B_\text{z}^\text{ext}$ is first ramped to a large value of $+(-) 300$ mT, respectively. In Fig.~\ref{fig:4}a, the slope of ESR transition $d (E_\text{Z} / h) / d B_\text{z}^\text{ext} = $ 44 MHz/mT, which is higher than a value of 28 MHz/mT expected from a fully saturated micromagnet. The slope of the transition suggests the micromagnet isn't fully polarized and has a magnetic susceptibility of $d B_\text{z}^\text{M} / d B_\text{z}^\text{ext} \approx 0.6$ around $B_\text{z}^\text{ext} = 92$ mT.

\textbf{Theory fits to $g_\text{s}$ and $\gamma_\text{s}$} Here we derive analytical expressions for the spin-photon coupling rate $g_\text{s}$ and spin dephasing rate $\gamma_\text{s}$. We focus on the $\epsilon = 0$ regime used in Fig.~\ref{fig:3}b. Accounting for spin-charge hybridization due to the field gradient $B_\text{x}^\text{M}$, the relevant DQD eigenstates are $\ket{0} \approx \ket{-, \downarrow}$, $\ket{1} \approx \cos{\frac{\Phi}{2}}\ket{-, \uparrow} + \sin{\frac{\Phi}{2}}\ket{+, \downarrow}$, $\ket{2} \approx \sin{\frac{\Phi}{2}}\ket{-, \uparrow} - \cos{\frac{\Phi}{2}}\ket{+, \downarrow}$ and $\ket{3} \approx \ket{+, \uparrow}$. Here we have introduced a mixing angle $\Phi = \tan^{-1} \frac{g \mu_\text{B} B_\text{x}^\text{M}}{2t_\text{c} - g \mu_\text{B} B_\text{z}}$, where $B_\text{z} = B_\text{z}^\text{ext} + B_\text{z}^\text{M}$ is the total magnetic field along the z-axis. The dipole transition matrix element for the primarily spin-like transition between $\ket{0}$ and $\ket{1}$ is given by $d_{01} \approx -\sin{\frac{\Phi}{2}}$, and the dipole transition matrix element for the primarily charge-like transition between $\ket{0}$ and $\ket{2}$ is given by $d_{02} \approx \cos{\frac{\Phi}{2}}$. The transition between $\ket{0}$ and $\ket{3}$ is too high in energy (off resonance) and is therefore excluded from our model. The spin-photon coupling rate is given by $g_\text{s} = g_\text{c} |d_{01}| = g_\text{c} |\sin{\frac{\Phi}{2}}|$, in agreement with previous theory works \cite{Hu_PRB_2012,Beaudoin_Nanotech_2017}.

To calculate the expected spin dephasing rate $\gamma_\text{s}^\text{(c)}$ arising from charge noise, we first construct the operators $\sigma_{01} = \ket{0} \bra{1} \approx \cos{\frac{\Phi}{2}} \sigma_\text{s} + \sin{\frac{\Phi}{2}} \sigma_\tau$ and $\sigma_{02} = \ket{0} \bra{2} \approx \sin{\frac{\Phi}{2}} \sigma_\text{s} - \cos{\frac{\Phi}{2}} \sigma_\tau$. Here $\sigma_\text{s} = \ket{-, \downarrow} \bra{-, \uparrow}$ and $\sigma_\tau = \ket{-, \downarrow} \bra{+, \downarrow}$ are lowering operators for the electron spin and charge respectively. Assuming the electron charge states dephase at a constant rate $\gamma_\text{c}$, the equations of motion for these operators are:

\begin{equation}
\dot{\sigma}_{01} = \gamma_\text{c} \left( -\sin^2{\frac{\Phi}{2}} \sigma_{01} + \frac{\sin{\Phi}}{2} \sigma_{02}  \right),
\end{equation}
\begin{equation}
\dot{\sigma}_{02} = \gamma_\text{c} \left( \frac{\sin{\Phi}}{2} \sigma_{01}  -\cos^2{\frac{\Phi}{2}} \sigma_{02}  \right).
\end{equation}

Combined with charge-photon coupling, the overall equations of motion in a rotating frame with a drive frequency $f \approx f_\text{c}$ are:

\begin{equation}
\dot{a} = i \Delta_0 a - \frac{\kappa}{2} a + \sqrt{\kappa_1}a_\text{in,1} - ig_\text{c} \left( d_{01} \sigma_{01} + d_{02} \sigma_{02} \right),
\end{equation}
\begin{equation}
\dot{\sigma}_{01} = -i\delta_1 \sigma_{01} -  \gamma_\text{c} \sin^2{\frac{\Phi}{2}} \sigma_{01} + \gamma_\text{c} \frac{\sin{\Phi}}{2} \sigma_{02} - ig_\text{c}ad_{10} , 
\end{equation}
\begin{equation}
\dot{\sigma}_{02} = -i\delta_2 \sigma_{02} -  \gamma_\text{c} \cos^2{\frac{\Phi}{2}} \sigma_{02} + \gamma_\text{c} \frac{\sin{\Phi}}{2} \sigma_{01} - ig_\text{c}ad_{20}  .
\end{equation}

Here $a$ is the photon annihilation operator of the cavity, $\Delta_0 = 2 \pi (f - f_\text{c})$ is the detuning between the cavity drive frequency from its center frequency, $\kappa_1$ is the photon decay rate at the input port of the cavity and $a_\text{in,1}$ is the input field of the cavity. The $\delta_1$ and $\delta_2$ terms are defined as $\delta_1 / 2\pi = (E_1 - E_0) / h - f$ and $\delta_2 / 2\pi = (E_2 - E_0) / h - f$, where $E_{0,1,2}$ corresponds to the energy of the $\ket{0}$, $\ket{1}$, or $\ket{2}$ state. Steady-state solutions to the above equations give the electric susceptibility of the spin qubit transition $\chi_\text{0,1} = \frac{\sigma_{01}}{a} = \frac{g_\text{s}}{\delta_1 - i \gamma_\text{s}^\text{(c)}}$, where we have identified a charge-induced spin dephasing rate $\gamma_\text{s}^\text{(c)} = \gamma_\text{c} \left[ \delta_2 \sin^2{\frac{\Phi}{2}} + \delta_1 \cos^2{\frac{\Phi}{2}} \right] / \delta_2$. To account for spin dephasing due to fluctuations of the $^{29}$Si nuclear spin bath, we express the total spin dephasing rate assuming a Voigt profile: $\gamma_\text{s} = \gamma_\text{s}^\text{(c)}/2 + \sqrt{(\gamma_\text{s}^\text{(c)}/2)^2 + 8 (\ln{2}) (1/T^*_\text{2,nuclear})^2}$, where $T^*_\text{2,nuclear}$  $\approx$ 1 $\mu$s is the electron spin dephasing time due to nuclear field fluctuations \cite{Kawakami.2014,Zajac_Arxiv_2017}.

In fitting to the data of Fig.~\ref{fig:3}b, we use the experimentally determined values of $g_\text{c} / 2 \pi =$ 40 MHz and $\gamma_\text{c} / 2 \pi =$ 35 MHz, along with a best fit field gradient $B_\text{x}^\text{M}$ = 15 mT. For every $t_\text{c}$, the fit value for $B_\text{z}$ is adjusted such that the spin qubit frequency $(E_1 - E_0) / h$ matches the cavity frequency $f_\text{c}$ exactly. The slight discrepancy between theory and experiment for $\gamma_\text{s}$ may be due to the frequency dependence of $\gamma_\text{c}$, changes in $\gamma_\text{c}$ with $B_\text{z}^{\rm ext}$, or other noise mechanisms not captured by this simple model. Detailed measurements of $\gamma_\text{c}$ and a comparison with a more sophisticated theoretical model will be the subject of future work.

\textbf{~\\Data Availability}\\
The data that support the findings of this study are available from the corresponding author on reasonable request.

\textbf{~\\Acknowledgements}\\
We thank A. J. Sigillito for technical assistance and M. J. Gullans for helpful discussions. Supported by the U.S. Department of Defense under contract H98230-15-C0453, Army Research Office grant W911NF-15-1-0149, and the Gordon and Betty Moore Foundations EPiQS Initiative through grant GBMF4535. Devices were fabricated in the Princeton University Quantum Device Nanofabrication Laboratory.

\textbf{~\\Author Contributions}\\
X.M. fabricated the sample and performed the measurements. X.M., D.M.Z. and J.R.P. developed the design and fabrication process for the DQD. X.M. and S.P. developed the Nb cavity fabrication process. M.B., G.B., J.M.T. and J.R.P. developed the theory for the experiment. X.M., M.B. and J.M.T. performed analysis of the data. X.M., J.R.P. and J.M.T. wrote the manuscript with input from the other authors. J.R.P. planned and supervised the experiment.

\textbf{~\\Additional Information}\\
Supplementary information is available in the online version of the paper. Reprints and permissions information is available online at www.nature.com/reprints.
Correspondence and requests for materials should be addressed to J.R.P. (petta@princeton.edu).

\pagebreak
\textbf{~\\Competing Financial Interests}\\
X.M., J.R.P., D.M.Z. and Princeton University have filed a provisional patent application related to spin-photon transduction. 

\bibliographystyle{apsrev4-1}
\bibliography{references_v8}

\begin{thebibliography}{45}%
\makeatletter
\providecommand \@ifxundefined [1]{%
 \@ifx{#1\undefined}
}%
\providecommand \@ifnum [1]{%
 \ifnum #1\expandafter \@firstoftwo
 \else \expandafter \@secondoftwo
 \fi
}%
\providecommand \@ifx [1]{%
 \ifx #1\expandafter \@firstoftwo
 \else \expandafter \@secondoftwo
 \fi
}%
\providecommand \natexlab [1]{#1}%
\providecommand \enquote  [1]{``#1''}%
\providecommand \bibnamefont  [1]{#1}%
\providecommand \bibfnamefont [1]{#1}%
\providecommand \citenamefont [1]{#1}%
\providecommand \href@noop [0]{\@secondoftwo}%
\providecommand \href [0]{\begingroup \@sanitize@url \@href}%
\providecommand \@href[1]{\@@startlink{#1}\@@href}%
\providecommand \@@href[1]{\endgroup#1\@@endlink}%
\providecommand \@sanitize@url [0]{\catcode `\\12\catcode `\$12\catcode
  `\&12\catcode `\#12\catcode `\^12\catcode `\_12\catcode `\%12\relax}%
\providecommand \@@startlink[1]{}%
\providecommand \@@endlink[0]{}%
\providecommand \url  [0]{\begingroup\@sanitize@url \@url }%
\providecommand \@url [1]{\endgroup\@href {#1}{\urlprefix }}%
\providecommand \urlprefix  [0]{URL }%
\providecommand \Eprint [0]{\href }%
\providecommand \doibase [0]{http://dx.doi.org/}%
\providecommand \selectlanguage [0]{\@gobble}%
\providecommand \bibinfo  [0]{\@secondoftwo}%
\providecommand \bibfield  [0]{\@secondoftwo}%
\providecommand \translation [1]{[#1]}%
\providecommand \BibitemOpen [0]{}%
\providecommand \bibitemStop [0]{}%
\providecommand \bibitemNoStop [0]{.\EOS\space}%
\providecommand \EOS [0]{\spacefactor3000\relax}%
\providecommand \BibitemShut  [1]{\csname bibitem#1\endcsname}%
\let\auto@bib@innerbib\@empty
\bibitem [{\citenamefont {Tyryshkin}\ \emph {et~al.}(2012)\citenamefont
  {Tyryshkin}, \citenamefont {Tojo}, \citenamefont {Morton}, \citenamefont
  {Riemann}, \citenamefont {Abrosimov}, \citenamefont {Becker}, \citenamefont
  {Pohl}, \citenamefont {Schenkel}, \citenamefont {Thewalt}, \citenamefont
  {Itoh},\ and\ \citenamefont {Lyon}}]{Lyon_NatMat_2012}%
  \BibitemOpen
  \bibfield  {author} {\bibinfo {author} {\bibfnamefont {A.~M.}\ \bibnamefont
  {Tyryshkin}}, \bibinfo {author} {\bibfnamefont {S.}~\bibnamefont {Tojo}},
  \bibinfo {author} {\bibfnamefont {J.~J.~L.}\ \bibnamefont {Morton}}, \bibinfo
  {author} {\bibfnamefont {H.}~\bibnamefont {Riemann}}, \bibinfo {author}
  {\bibfnamefont {N.~V.}\ \bibnamefont {Abrosimov}}, \bibinfo {author}
  {\bibfnamefont {P.}~\bibnamefont {Becker}}, \bibinfo {author} {\bibfnamefont
  {H.-J.}\ \bibnamefont {Pohl}}, \bibinfo {author} {\bibfnamefont
  {T.}~\bibnamefont {Schenkel}}, \bibinfo {author} {\bibfnamefont {M.~L.~W.}\
  \bibnamefont {Thewalt}}, \bibinfo {author} {\bibfnamefont {K.~M.}\
  \bibnamefont {Itoh}}, \ and\ \bibinfo {author} {\bibfnamefont {S.~A.}\
  \bibnamefont {Lyon}},\ }\href@noop {} {\bibfield  {journal} {\bibinfo
  {journal} {Nat. Mater.}\ }\textbf {\bibinfo {volume} {11}},\ \bibinfo {pages}
  {143} (\bibinfo {year} {2012})}\BibitemShut {NoStop}%
\bibitem [{\citenamefont {Saeedi}\ \emph {et~al.}(2013)\citenamefont {Saeedi},
  \citenamefont {Simmons}, \citenamefont {Salvail}, \citenamefont {Dluhy},
  \citenamefont {Riemann}, \citenamefont {Abrosimov}, \citenamefont {Becker},
  \citenamefont {Pohl}, \citenamefont {Morton},\ and\ \citenamefont
  {Thewalt}}]{Saeedi2013}%
  \BibitemOpen
  \bibfield  {author} {\bibinfo {author} {\bibfnamefont {K.}~\bibnamefont
  {Saeedi}}, \bibinfo {author} {\bibfnamefont {S.}~\bibnamefont {Simmons}},
  \bibinfo {author} {\bibfnamefont {J.~Z.}\ \bibnamefont {Salvail}}, \bibinfo
  {author} {\bibfnamefont {P.}~\bibnamefont {Dluhy}}, \bibinfo {author}
  {\bibfnamefont {H.}~\bibnamefont {Riemann}}, \bibinfo {author} {\bibfnamefont
  {N.~V.}\ \bibnamefont {Abrosimov}}, \bibinfo {author} {\bibfnamefont
  {P.}~\bibnamefont {Becker}}, \bibinfo {author} {\bibfnamefont {H.-J.}\
  \bibnamefont {Pohl}}, \bibinfo {author} {\bibfnamefont {J.~J.~L.}\
  \bibnamefont {Morton}}, \ and\ \bibinfo {author} {\bibfnamefont {M.~L.~W.}\
  \bibnamefont {Thewalt}},\ }\href@noop {} {\bibfield  {journal} {\bibinfo
  {journal} {Science}\ }\textbf {\bibinfo {volume} {342}},\ \bibinfo {pages}
  {830} (\bibinfo {year} {2013})}\BibitemShut {NoStop}%
\bibitem [{\citenamefont {Petta}\ \emph {et~al.}(2005)\citenamefont {Petta},
  \citenamefont {Johnson}, \citenamefont {Taylor}, \citenamefont {Laird},
  \citenamefont {Yacoby}, \citenamefont {Lukin}, \citenamefont {Marcus},
  \citenamefont {Hanson},\ and\ \citenamefont {Gossard}}]{Petta_Science}%
  \BibitemOpen
  \bibfield  {author} {\bibinfo {author} {\bibfnamefont {J.~R.}\ \bibnamefont
  {Petta}}, \bibinfo {author} {\bibfnamefont {A.~C.}\ \bibnamefont {Johnson}},
  \bibinfo {author} {\bibfnamefont {J.~M.}\ \bibnamefont {Taylor}}, \bibinfo
  {author} {\bibfnamefont {E.~A.}\ \bibnamefont {Laird}}, \bibinfo {author}
  {\bibfnamefont {A.}~\bibnamefont {Yacoby}}, \bibinfo {author} {\bibfnamefont
  {M.~D.}\ \bibnamefont {Lukin}}, \bibinfo {author} {\bibfnamefont {C.~M.}\
  \bibnamefont {Marcus}}, \bibinfo {author} {\bibfnamefont {M.~P.}\
  \bibnamefont {Hanson}}, \ and\ \bibinfo {author} {\bibfnamefont {A.~C.}\
  \bibnamefont {Gossard}},\ }\href@noop {} {\bibfield  {journal} {\bibinfo
  {journal} {Science}\ }\textbf {\bibinfo {volume} {309}},\ \bibinfo {pages}
  {2180} (\bibinfo {year} {2005})}\BibitemShut {NoStop}%
\bibitem [{\citenamefont {Neumann}\ \emph {et~al.}(2008)\citenamefont
  {Neumann}, \citenamefont {Mizuochi}, \citenamefont {Rempp}, \citenamefont
  {Hemmer}, \citenamefont {Watanabe}, \citenamefont {Yamasaki}, \citenamefont
  {Jacques}, \citenamefont {Gaebel}, \citenamefont {Jelezko},\ and\
  \citenamefont {Wrachtrup}}]{Neumann1326}%
  \BibitemOpen
  \bibfield  {author} {\bibinfo {author} {\bibfnamefont {P.}~\bibnamefont
  {Neumann}}, \bibinfo {author} {\bibfnamefont {N.}~\bibnamefont {Mizuochi}},
  \bibinfo {author} {\bibfnamefont {F.}~\bibnamefont {Rempp}}, \bibinfo
  {author} {\bibfnamefont {P.}~\bibnamefont {Hemmer}}, \bibinfo {author}
  {\bibfnamefont {H.}~\bibnamefont {Watanabe}}, \bibinfo {author}
  {\bibfnamefont {S.}~\bibnamefont {Yamasaki}}, \bibinfo {author}
  {\bibfnamefont {V.}~\bibnamefont {Jacques}}, \bibinfo {author} {\bibfnamefont
  {T.}~\bibnamefont {Gaebel}}, \bibinfo {author} {\bibfnamefont
  {F.}~\bibnamefont {Jelezko}}, \ and\ \bibinfo {author} {\bibfnamefont
  {J.}~\bibnamefont {Wrachtrup}},\ }\href@noop {} {\bibfield  {journal}
  {\bibinfo  {journal} {Science}\ }\textbf {\bibinfo {volume} {320}},\ \bibinfo
  {pages} {1326} (\bibinfo {year} {2008})}\BibitemShut {NoStop}%
\bibitem [{\citenamefont {Dehollain}\ \emph {et~al.}(2016)\citenamefont
  {Dehollain}, \citenamefont {Simmons}, \citenamefont {Muhonen}, \citenamefont
  {Kalra}, \citenamefont {Laucht}, \citenamefont {Hudson}, \citenamefont
  {Itoh}, \citenamefont {Jamieson}, \citenamefont {McCallum}, \citenamefont
  {Dzurak},\ and\ \citenamefont {Morello}}]{Dehollain2016}%
  \BibitemOpen
  \bibfield  {author} {\bibinfo {author} {\bibfnamefont {J.~P.}\ \bibnamefont
  {Dehollain}}, \bibinfo {author} {\bibfnamefont {S.}~\bibnamefont {Simmons}},
  \bibinfo {author} {\bibfnamefont {J.~T.}\ \bibnamefont {Muhonen}}, \bibinfo
  {author} {\bibfnamefont {R.}~\bibnamefont {Kalra}}, \bibinfo {author}
  {\bibfnamefont {A.}~\bibnamefont {Laucht}}, \bibinfo {author} {\bibfnamefont
  {F.}~\bibnamefont {Hudson}}, \bibinfo {author} {\bibfnamefont {K.~M.}\
  \bibnamefont {Itoh}}, \bibinfo {author} {\bibfnamefont {D.~N.}\ \bibnamefont
  {Jamieson}}, \bibinfo {author} {\bibfnamefont {J.~C.}\ \bibnamefont
  {McCallum}}, \bibinfo {author} {\bibfnamefont {A.~S.}\ \bibnamefont
  {Dzurak}}, \ and\ \bibinfo {author} {\bibfnamefont {A.}~\bibnamefont
  {Morello}},\ }\href@noop {} {\bibfield  {journal} {\bibinfo  {journal} {Nat.
  Nanotechnol.}\ }\textbf {\bibinfo {volume} {11}},\ \bibinfo {pages} {242}
  (\bibinfo {year} {2016})}\BibitemShut {NoStop}%
\bibitem [{\citenamefont {Hanson}\ \emph {et~al.}(2007)\citenamefont {Hanson},
  \citenamefont {Kouwenhoven}, \citenamefont {Petta}, \citenamefont {Tarucha},\
  and\ \citenamefont {Vandersypen}}]{Petta_RevMod}%
  \BibitemOpen
  \bibfield  {author} {\bibinfo {author} {\bibfnamefont {R.}~\bibnamefont
  {Hanson}}, \bibinfo {author} {\bibfnamefont {L.~P.}\ \bibnamefont
  {Kouwenhoven}}, \bibinfo {author} {\bibfnamefont {J.~R.}\ \bibnamefont
  {Petta}}, \bibinfo {author} {\bibfnamefont {S.}~\bibnamefont {Tarucha}}, \
  and\ \bibinfo {author} {\bibfnamefont {L.~M.~K.}\ \bibnamefont
  {Vandersypen}},\ }\href@noop {} {\bibfield  {journal} {\bibinfo  {journal}
  {Rev. Mod. Phys.}\ }\textbf {\bibinfo {volume} {79}},\ \bibinfo {pages}
  {1217} (\bibinfo {year} {2007})}\BibitemShut {NoStop}%
\bibitem [{\citenamefont {Zwanenburg}\ \emph {et~al.}(2013)\citenamefont
  {Zwanenburg}, \citenamefont {Dzurak}, \citenamefont {Morello}, \citenamefont
  {Simmons}, \citenamefont {Hollenberg}, \citenamefont {Klimeck}, \citenamefont
  {Rogge}, \citenamefont {Coppersmith},\ and\ \citenamefont
  {Eriksson}}]{RevModPhys.85.961}%
  \BibitemOpen
  \bibfield  {author} {\bibinfo {author} {\bibfnamefont {F.~A.}\ \bibnamefont
  {Zwanenburg}}, \bibinfo {author} {\bibfnamefont {A.~S.}\ \bibnamefont
  {Dzurak}}, \bibinfo {author} {\bibfnamefont {A.}~\bibnamefont {Morello}},
  \bibinfo {author} {\bibfnamefont {M.~Y.}\ \bibnamefont {Simmons}}, \bibinfo
  {author} {\bibfnamefont {L.~C.~L.}\ \bibnamefont {Hollenberg}}, \bibinfo
  {author} {\bibfnamefont {G.}~\bibnamefont {Klimeck}}, \bibinfo {author}
  {\bibfnamefont {S.}~\bibnamefont {Rogge}}, \bibinfo {author} {\bibfnamefont
  {S.~N.}\ \bibnamefont {Coppersmith}}, \ and\ \bibinfo {author} {\bibfnamefont
  {M.~A.}\ \bibnamefont {Eriksson}},\ }\href@noop {} {\bibfield  {journal}
  {\bibinfo  {journal} {Rev. Mod. Phys.}\ }\textbf {\bibinfo {volume} {85}},\
  \bibinfo {pages} {961} (\bibinfo {year} {2013})}\BibitemShut {NoStop}%
\bibitem [{\citenamefont {Veldhorst}\ \emph {et~al.}(2014)\citenamefont
  {Veldhorst}, \citenamefont {Hwang}, \citenamefont {Yang}, \citenamefont
  {Leenstra}, \citenamefont {de~Ronde}, \citenamefont {Dehollain},
  \citenamefont {Muhonen}, \citenamefont {Hudson}, \citenamefont {Itoh},
  \citenamefont {Morello},\ and\ \citenamefont {Dzurak}}]{VeldhorstM.2014}%
  \BibitemOpen
  \bibfield  {author} {\bibinfo {author} {\bibfnamefont {M.}~\bibnamefont
  {Veldhorst}}, \bibinfo {author} {\bibfnamefont {J.~C.~C.}\ \bibnamefont
  {Hwang}}, \bibinfo {author} {\bibfnamefont {C.~H.}\ \bibnamefont {Yang}},
  \bibinfo {author} {\bibfnamefont {A.~W.}\ \bibnamefont {Leenstra}}, \bibinfo
  {author} {\bibfnamefont {B.}~\bibnamefont {de~Ronde}}, \bibinfo {author}
  {\bibfnamefont {J.~P.}\ \bibnamefont {Dehollain}}, \bibinfo {author}
  {\bibfnamefont {J.~T.}\ \bibnamefont {Muhonen}}, \bibinfo {author}
  {\bibfnamefont {F.~E.}\ \bibnamefont {Hudson}}, \bibinfo {author}
  {\bibfnamefont {K.~M.}\ \bibnamefont {Itoh}}, \bibinfo {author}
  {\bibfnamefont {A.}~\bibnamefont {Morello}}, \ and\ \bibinfo {author}
  {\bibfnamefont {A.~S.}\ \bibnamefont {Dzurak}},\ }\href@noop {} {\bibfield
  {journal} {\bibinfo  {journal} {Nat. Nanotechnol.}\ }\textbf {\bibinfo
  {volume} {9}},\ \bibinfo {pages} {981} (\bibinfo {year} {2014})}\BibitemShut
  {NoStop}%
\bibitem [{\citenamefont {Takeda}\ \emph {et~al.}(2016)\citenamefont {Takeda},
  \citenamefont {Kamioka}, \citenamefont {Otsuka}, \citenamefont {Yoneda},
  \citenamefont {Nakajima}, \citenamefont {Delbecq}, \citenamefont {Amaha},
  \citenamefont {Allison}, \citenamefont {Kodera}, \citenamefont {Oda},\ and\
  \citenamefont {Tarucha}}]{Takeda2016}%
  \BibitemOpen
  \bibfield  {author} {\bibinfo {author} {\bibfnamefont {K.}~\bibnamefont
  {Takeda}}, \bibinfo {author} {\bibfnamefont {J.}~\bibnamefont {Kamioka}},
  \bibinfo {author} {\bibfnamefont {T.}~\bibnamefont {Otsuka}}, \bibinfo
  {author} {\bibfnamefont {J.}~\bibnamefont {Yoneda}}, \bibinfo {author}
  {\bibfnamefont {T.}~\bibnamefont {Nakajima}}, \bibinfo {author}
  {\bibfnamefont {M.~R.}\ \bibnamefont {Delbecq}}, \bibinfo {author}
  {\bibfnamefont {S.}~\bibnamefont {Amaha}}, \bibinfo {author} {\bibfnamefont
  {G.}~\bibnamefont {Allison}}, \bibinfo {author} {\bibfnamefont
  {T.}~\bibnamefont {Kodera}}, \bibinfo {author} {\bibfnamefont
  {S.}~\bibnamefont {Oda}}, \ and\ \bibinfo {author} {\bibfnamefont
  {S.}~\bibnamefont {Tarucha}},\ }\href@noop {} {\bibfield  {journal} {\bibinfo
   {journal} {Sci. Adv.}\ }\textbf {\bibinfo {volume} {2}} (\bibinfo {year}
  {2016})}\BibitemShut {NoStop}%
\bibitem [{\citenamefont {Veldhorst}\ \emph {et~al.}(2015)\citenamefont
  {Veldhorst}, \citenamefont {Yang}, \citenamefont {Hwang}, \citenamefont
  {Huang}, \citenamefont {Dehollain}, \citenamefont {Muhonen}, \citenamefont
  {Simmons}, \citenamefont {Laucht}, \citenamefont {Hudson}, \citenamefont
  {Itoh}, \citenamefont {Morello},\ and\ \citenamefont
  {Dzurak}}]{Veldhorst2015}%
  \BibitemOpen
  \bibfield  {author} {\bibinfo {author} {\bibfnamefont {M.}~\bibnamefont
  {Veldhorst}}, \bibinfo {author} {\bibfnamefont {C.~H.}\ \bibnamefont {Yang}},
  \bibinfo {author} {\bibfnamefont {J.~C.~C.}\ \bibnamefont {Hwang}}, \bibinfo
  {author} {\bibfnamefont {W.}~\bibnamefont {Huang}}, \bibinfo {author}
  {\bibfnamefont {J.~P.}\ \bibnamefont {Dehollain}}, \bibinfo {author}
  {\bibfnamefont {J.~T.}\ \bibnamefont {Muhonen}}, \bibinfo {author}
  {\bibfnamefont {S.}~\bibnamefont {Simmons}}, \bibinfo {author} {\bibfnamefont
  {A.}~\bibnamefont {Laucht}}, \bibinfo {author} {\bibfnamefont {F.~E.}\
  \bibnamefont {Hudson}}, \bibinfo {author} {\bibfnamefont {K.~M.}\
  \bibnamefont {Itoh}}, \bibinfo {author} {\bibfnamefont {A.}~\bibnamefont
  {Morello}}, \ and\ \bibinfo {author} {\bibfnamefont {A.~S.}\ \bibnamefont
  {Dzurak}},\ }\href@noop {} {\bibfield  {journal} {\bibinfo  {journal}
  {Nature}\ }\textbf {\bibinfo {volume} {526}},\ \bibinfo {pages} {410}
  (\bibinfo {year} {2015})}\BibitemShut {NoStop}%
\bibitem [{\citenamefont {{Zajac}}\ \emph {et~al.}(2017)\citenamefont
  {{Zajac}}, \citenamefont {{Sigillito}}, \citenamefont {{Russ}}, \citenamefont
  {{Borjans}}, \citenamefont {{Taylor}}, \citenamefont {{Burkard}},\ and\
  \citenamefont {{Petta}}}]{Zajac_Arxiv_2017}%
  \BibitemOpen
  \bibfield  {author} {\bibinfo {author} {\bibfnamefont {D.~M.}\ \bibnamefont
  {{Zajac}}}, \bibinfo {author} {\bibfnamefont {A.~J.}\ \bibnamefont
  {{Sigillito}}}, \bibinfo {author} {\bibfnamefont {M.}~\bibnamefont {{Russ}}},
  \bibinfo {author} {\bibfnamefont {F.}~\bibnamefont {{Borjans}}}, \bibinfo
  {author} {\bibfnamefont {J.~M.}\ \bibnamefont {{Taylor}}}, \bibinfo {author}
  {\bibfnamefont {G.}~\bibnamefont {{Burkard}}}, \ and\ \bibinfo {author}
  {\bibfnamefont {J.~R.}\ \bibnamefont {{Petta}}},\ }\href@noop {} {\bibfield
  {journal} {\bibinfo  {journal} {arXiv:1708.03530}\ } (\bibinfo {year}
  {2017})}\BibitemShut {NoStop}%
\bibitem [{\citenamefont {{Watson}}\ \emph {et~al.}(2017)\citenamefont
  {{Watson}}, \citenamefont {{Philips}}, \citenamefont {{Kawakami}},
  \citenamefont {{Ward}}, \citenamefont {{Scarlino}}, \citenamefont
  {{Veldhorst}}, \citenamefont {{Savage}}, \citenamefont {{Lagally}},
  \citenamefont {{Friesen}}, \citenamefont {{Coppersmith}}, \citenamefont
  {{Eriksson}},\ and\ \citenamefont {{Vandersypen}}}]{Watson_Arxiv_2017}%
  \BibitemOpen
  \bibfield  {author} {\bibinfo {author} {\bibfnamefont {T.~F.}\ \bibnamefont
  {{Watson}}}, \bibinfo {author} {\bibfnamefont {S.~G.~J.}\ \bibnamefont
  {{Philips}}}, \bibinfo {author} {\bibfnamefont {E.}~\bibnamefont
  {{Kawakami}}}, \bibinfo {author} {\bibfnamefont {D.~R.}\ \bibnamefont
  {{Ward}}}, \bibinfo {author} {\bibfnamefont {P.}~\bibnamefont {{Scarlino}}},
  \bibinfo {author} {\bibfnamefont {M.}~\bibnamefont {{Veldhorst}}}, \bibinfo
  {author} {\bibfnamefont {D.~E.}\ \bibnamefont {{Savage}}}, \bibinfo {author}
  {\bibfnamefont {M.~G.}\ \bibnamefont {{Lagally}}}, \bibinfo {author}
  {\bibfnamefont {M.}~\bibnamefont {{Friesen}}}, \bibinfo {author}
  {\bibfnamefont {S.~N.}\ \bibnamefont {{Coppersmith}}}, \bibinfo {author}
  {\bibfnamefont {M.~A.}\ \bibnamefont {{Eriksson}}}, \ and\ \bibinfo {author}
  {\bibfnamefont {L.~M.~K.}\ \bibnamefont {{Vandersypen}}},\ }\href@noop {}
  {\bibfield  {journal} {\bibinfo  {journal} {arXiv:1708.04214}\ } (\bibinfo
  {year} {2017})}\BibitemShut {NoStop}%
\bibitem [{\citenamefont {Mi}\ \emph {et~al.}(2017{\natexlab{a}})\citenamefont
  {Mi}, \citenamefont {Cady}, \citenamefont {Zajac}, \citenamefont {Deelman},\
  and\ \citenamefont {Petta}}]{Mi_Science_2016}%
  \BibitemOpen
  \bibfield  {author} {\bibinfo {author} {\bibfnamefont {X.}~\bibnamefont
  {Mi}}, \bibinfo {author} {\bibfnamefont {J.~V.}\ \bibnamefont {Cady}},
  \bibinfo {author} {\bibfnamefont {D.~M.}\ \bibnamefont {Zajac}}, \bibinfo
  {author} {\bibfnamefont {P.~W.}\ \bibnamefont {Deelman}}, \ and\ \bibinfo
  {author} {\bibfnamefont {J.~R.}\ \bibnamefont {Petta}},\ }\href@noop {}
  {\bibfield  {journal} {\bibinfo  {journal} {Science}\ }\textbf {\bibinfo
  {volume} {355}},\ \bibinfo {pages} {156} (\bibinfo {year}
  {2017}{\natexlab{a}})}\BibitemShut {NoStop}%
\bibitem [{\citenamefont {Thompson}\ \emph {et~al.}(1992)\citenamefont
  {Thompson}, \citenamefont {Rempe},\ and\ \citenamefont
  {Kimble}}]{PhysRevLett.68.1132}%
  \BibitemOpen
  \bibfield  {author} {\bibinfo {author} {\bibfnamefont {R.~J.}\ \bibnamefont
  {Thompson}}, \bibinfo {author} {\bibfnamefont {G.}~\bibnamefont {Rempe}}, \
  and\ \bibinfo {author} {\bibfnamefont {H.~J.}\ \bibnamefont {Kimble}},\
  }\href@noop {} {\bibfield  {journal} {\bibinfo  {journal} {Phys. Rev. Lett.}\
  }\textbf {\bibinfo {volume} {68}},\ \bibinfo {pages} {1132} (\bibinfo {year}
  {1992})}\BibitemShut {NoStop}%
\bibitem [{\citenamefont {Brune}\ \emph {et~al.}(1996)\citenamefont {Brune},
  \citenamefont {Schmidt-Kaler}, \citenamefont {Maali}, \citenamefont {Dreyer},
  \citenamefont {Hagley}, \citenamefont {Raimond},\ and\ \citenamefont
  {Haroche}}]{PhysRevLett.76.1800}%
  \BibitemOpen
  \bibfield  {author} {\bibinfo {author} {\bibfnamefont {M.}~\bibnamefont
  {Brune}}, \bibinfo {author} {\bibfnamefont {F.}~\bibnamefont
  {Schmidt-Kaler}}, \bibinfo {author} {\bibfnamefont {A.}~\bibnamefont
  {Maali}}, \bibinfo {author} {\bibfnamefont {J.}~\bibnamefont {Dreyer}},
  \bibinfo {author} {\bibfnamefont {E.}~\bibnamefont {Hagley}}, \bibinfo
  {author} {\bibfnamefont {J.~M.}\ \bibnamefont {Raimond}}, \ and\ \bibinfo
  {author} {\bibfnamefont {S.}~\bibnamefont {Haroche}},\ }\href@noop {}
  {\bibfield  {journal} {\bibinfo  {journal} {Phys. Rev. Lett.}\ }\textbf
  {\bibinfo {volume} {76}},\ \bibinfo {pages} {1800} (\bibinfo {year}
  {1996})}\BibitemShut {NoStop}%
\bibitem [{\citenamefont {Wallraff}\ \emph {et~al.}(2004)\citenamefont
  {Wallraff}, \citenamefont {Schuster}, \citenamefont {Blais}, \citenamefont
  {Frunzio}, \citenamefont {Huang}, \citenamefont {Majer}, \citenamefont
  {Kumar}, \citenamefont {Girvin},\ and\ \citenamefont
  {Schoelkopf}}]{Strong_Coupling_CooperPair}%
  \BibitemOpen
  \bibfield  {author} {\bibinfo {author} {\bibfnamefont {A.}~\bibnamefont
  {Wallraff}}, \bibinfo {author} {\bibfnamefont {D.~I.}\ \bibnamefont
  {Schuster}}, \bibinfo {author} {\bibfnamefont {A.}~\bibnamefont {Blais}},
  \bibinfo {author} {\bibfnamefont {L.}~\bibnamefont {Frunzio}}, \bibinfo
  {author} {\bibfnamefont {R.-S.}\ \bibnamefont {Huang}}, \bibinfo {author}
  {\bibfnamefont {J.}~\bibnamefont {Majer}}, \bibinfo {author} {\bibfnamefont
  {S.}~\bibnamefont {Kumar}}, \bibinfo {author} {\bibfnamefont {S.~M.}\
  \bibnamefont {Girvin}}, \ and\ \bibinfo {author} {\bibfnamefont {R.~J.}\
  \bibnamefont {Schoelkopf}},\ }\href@noop {} {\bibfield  {journal} {\bibinfo
  {journal} {Nature}\ }\textbf {\bibinfo {volume} {431}},\ \bibinfo {pages}
  {162} (\bibinfo {year} {2004})}\BibitemShut {NoStop}%
\bibitem [{\citenamefont {Childress}\ \emph {et~al.}(2004)\citenamefont
  {Childress}, \citenamefont {S\o{}rensen},\ and\ \citenamefont
  {Lukin}}]{PhysRevA.69.042302}%
  \BibitemOpen
  \bibfield  {author} {\bibinfo {author} {\bibfnamefont {L.}~\bibnamefont
  {Childress}}, \bibinfo {author} {\bibfnamefont {A.~S.}\ \bibnamefont
  {S\o{}rensen}}, \ and\ \bibinfo {author} {\bibfnamefont {M.~D.}\ \bibnamefont
  {Lukin}},\ }\href@noop {} {\bibfield  {journal} {\bibinfo  {journal} {Phys.
  Rev. A}\ }\textbf {\bibinfo {volume} {69}},\ \bibinfo {pages} {042302}
  (\bibinfo {year} {2004})}\BibitemShut {NoStop}%
\bibitem [{\citenamefont {Stockklauser}\ \emph {et~al.}(2017)\citenamefont
  {Stockklauser}, \citenamefont {Scarlino}, \citenamefont {Koski},
  \citenamefont {Gasparinetti}, \citenamefont {Andersen}, \citenamefont
  {Reichl}, \citenamefont {Wegscheider}, \citenamefont {Ihn}, \citenamefont
  {Ensslin},\ and\ \citenamefont {Wallraff}}]{PhysRevX.7.011030}%
  \BibitemOpen
  \bibfield  {author} {\bibinfo {author} {\bibfnamefont {A.}~\bibnamefont
  {Stockklauser}}, \bibinfo {author} {\bibfnamefont {P.}~\bibnamefont
  {Scarlino}}, \bibinfo {author} {\bibfnamefont {J.~V.}\ \bibnamefont {Koski}},
  \bibinfo {author} {\bibfnamefont {S.}~\bibnamefont {Gasparinetti}}, \bibinfo
  {author} {\bibfnamefont {C.~K.}\ \bibnamefont {Andersen}}, \bibinfo {author}
  {\bibfnamefont {C.}~\bibnamefont {Reichl}}, \bibinfo {author} {\bibfnamefont
  {W.}~\bibnamefont {Wegscheider}}, \bibinfo {author} {\bibfnamefont
  {T.}~\bibnamefont {Ihn}}, \bibinfo {author} {\bibfnamefont {K.}~\bibnamefont
  {Ensslin}}, \ and\ \bibinfo {author} {\bibfnamefont {A.}~\bibnamefont
  {Wallraff}},\ }\href@noop {} {\bibfield  {journal} {\bibinfo  {journal}
  {Phys. Rev. X}\ }\textbf {\bibinfo {volume} {7}},\ \bibinfo {pages} {011030}
  (\bibinfo {year} {2017})}\BibitemShut {NoStop}%
\bibitem [{\citenamefont {Imamo\ifmmode~\breve{g}\else
  \u{g}\fi{}lu}(2009)}]{PhysRevLett.102.083602}%
  \BibitemOpen
  \bibfield  {author} {\bibinfo {author} {\bibfnamefont {A.}~\bibnamefont
  {Imamo\ifmmode~\breve{g}\else \u{g}\fi{}lu}},\ }\href@noop {} {\bibfield
  {journal} {\bibinfo  {journal} {Phys. Rev. Lett.}\ }\textbf {\bibinfo
  {volume} {102}},\ \bibinfo {pages} {083602} (\bibinfo {year}
  {2009})}\BibitemShut {NoStop}%
\bibitem [{\citenamefont {Schuster}\ \emph {et~al.}(2010)\citenamefont
  {Schuster}, \citenamefont {Sears}, \citenamefont {Ginossar}, \citenamefont
  {DiCarlo}, \citenamefont {Frunzio}, \citenamefont {Morton}, \citenamefont
  {Wu}, \citenamefont {Briggs}, \citenamefont {Buckley}, \citenamefont
  {Awschalom},\ and\ \citenamefont {Schoelkopf}}]{PhysRevLett.105.140501}%
  \BibitemOpen
  \bibfield  {author} {\bibinfo {author} {\bibfnamefont {D.~I.}\ \bibnamefont
  {Schuster}}, \bibinfo {author} {\bibfnamefont {A.~P.}\ \bibnamefont {Sears}},
  \bibinfo {author} {\bibfnamefont {E.}~\bibnamefont {Ginossar}}, \bibinfo
  {author} {\bibfnamefont {L.}~\bibnamefont {DiCarlo}}, \bibinfo {author}
  {\bibfnamefont {L.}~\bibnamefont {Frunzio}}, \bibinfo {author} {\bibfnamefont
  {J.~J.~L.}\ \bibnamefont {Morton}}, \bibinfo {author} {\bibfnamefont
  {H.}~\bibnamefont {Wu}}, \bibinfo {author} {\bibfnamefont {G.~A.~D.}\
  \bibnamefont {Briggs}}, \bibinfo {author} {\bibfnamefont {B.~B.}\
  \bibnamefont {Buckley}}, \bibinfo {author} {\bibfnamefont {D.~D.}\
  \bibnamefont {Awschalom}}, \ and\ \bibinfo {author} {\bibfnamefont {R.~J.}\
  \bibnamefont {Schoelkopf}},\ }\href@noop {} {\bibfield  {journal} {\bibinfo
  {journal} {Phys. Rev. Lett.}\ }\textbf {\bibinfo {volume} {105}},\ \bibinfo
  {pages} {140501} (\bibinfo {year} {2010})}\BibitemShut {NoStop}%
\bibitem [{\citenamefont {Ams\"uss}\ \emph {et~al.}(2011)\citenamefont
  {Ams\"uss}, \citenamefont {Koller}, \citenamefont {N\"obauer}, \citenamefont
  {Putz}, \citenamefont {Rotter}, \citenamefont {Sandner}, \citenamefont
  {Schneider}, \citenamefont {Schramb\"ock}, \citenamefont {Steinhauser},
  \citenamefont {Ritsch}, \citenamefont {Schmiedmayer},\ and\ \citenamefont
  {Majer}}]{PhysRevLett.107.060502}%
  \BibitemOpen
  \bibfield  {author} {\bibinfo {author} {\bibfnamefont {R.}~\bibnamefont
  {Ams\"uss}}, \bibinfo {author} {\bibfnamefont {C.}~\bibnamefont {Koller}},
  \bibinfo {author} {\bibfnamefont {T.}~\bibnamefont {N\"obauer}}, \bibinfo
  {author} {\bibfnamefont {S.}~\bibnamefont {Putz}}, \bibinfo {author}
  {\bibfnamefont {S.}~\bibnamefont {Rotter}}, \bibinfo {author} {\bibfnamefont
  {K.}~\bibnamefont {Sandner}}, \bibinfo {author} {\bibfnamefont
  {S.}~\bibnamefont {Schneider}}, \bibinfo {author} {\bibfnamefont
  {M.}~\bibnamefont {Schramb\"ock}}, \bibinfo {author} {\bibfnamefont
  {G.}~\bibnamefont {Steinhauser}}, \bibinfo {author} {\bibfnamefont
  {H.}~\bibnamefont {Ritsch}}, \bibinfo {author} {\bibfnamefont
  {J.}~\bibnamefont {Schmiedmayer}}, \ and\ \bibinfo {author} {\bibfnamefont
  {J.}~\bibnamefont {Majer}},\ }\href@noop {} {\bibfield  {journal} {\bibinfo
  {journal} {Phys. Rev. Lett.}\ }\textbf {\bibinfo {volume} {107}},\ \bibinfo
  {pages} {060502} (\bibinfo {year} {2011})}\BibitemShut {NoStop}%
\bibitem [{\citenamefont {Bienfait}\ \emph {et~al.}(2016)\citenamefont
  {Bienfait}, \citenamefont {Pla}, \citenamefont {Kubo}, \citenamefont {Zhou},
  \citenamefont {Stern}, \citenamefont {Lo}, \citenamefont {Weis},
  \citenamefont {Schenkel}, \citenamefont {Vion}, \citenamefont {Esteve},
  \citenamefont {Morton},\ and\ \citenamefont {Bertet}}]{Bienfait_Nature_2016}%
  \BibitemOpen
  \bibfield  {author} {\bibinfo {author} {\bibfnamefont {A.}~\bibnamefont
  {Bienfait}}, \bibinfo {author} {\bibfnamefont {J.~J.}\ \bibnamefont {Pla}},
  \bibinfo {author} {\bibfnamefont {Y.}~\bibnamefont {Kubo}}, \bibinfo {author}
  {\bibfnamefont {X.}~\bibnamefont {Zhou}}, \bibinfo {author} {\bibfnamefont
  {M.}~\bibnamefont {Stern}}, \bibinfo {author} {\bibfnamefont {C.~C.}\
  \bibnamefont {Lo}}, \bibinfo {author} {\bibfnamefont {C.~D.}\ \bibnamefont
  {Weis}}, \bibinfo {author} {\bibfnamefont {T.}~\bibnamefont {Schenkel}},
  \bibinfo {author} {\bibfnamefont {D.}~\bibnamefont {Vion}}, \bibinfo {author}
  {\bibfnamefont {D.}~\bibnamefont {Esteve}}, \bibinfo {author} {\bibfnamefont
  {J.~J.~L.}\ \bibnamefont {Morton}}, \ and\ \bibinfo {author} {\bibfnamefont
  {P.}~\bibnamefont {Bertet}},\ }\href@noop {} {\bibfield  {journal} {\bibinfo
  {journal} {Nature}\ }\textbf {\bibinfo {volume} {531}},\ \bibinfo {pages}
  {74} (\bibinfo {year} {2016})}\BibitemShut {NoStop}%
\bibitem [{\citenamefont {Eichler}\ \emph {et~al.}(2017)\citenamefont
  {Eichler}, \citenamefont {Sigillito}, \citenamefont {Lyon},\ and\
  \citenamefont {Petta}}]{PhysRevLett.118.037701}%
  \BibitemOpen
  \bibfield  {author} {\bibinfo {author} {\bibfnamefont {C.}~\bibnamefont
  {Eichler}}, \bibinfo {author} {\bibfnamefont {A.~J.}\ \bibnamefont
  {Sigillito}}, \bibinfo {author} {\bibfnamefont {S.~A.}\ \bibnamefont {Lyon}},
  \ and\ \bibinfo {author} {\bibfnamefont {J.~R.}\ \bibnamefont {Petta}},\
  }\href@noop {} {\bibfield  {journal} {\bibinfo  {journal} {Phys. Rev. Lett.}\
  }\textbf {\bibinfo {volume} {118}},\ \bibinfo {pages} {037701} (\bibinfo
  {year} {2017})}\BibitemShut {NoStop}%
\bibitem [{\citenamefont {Trif}\ \emph {et~al.}(2008)\citenamefont {Trif},
  \citenamefont {Golovach},\ and\ \citenamefont {Loss}}]{Trif_PRB_2008}%
  \BibitemOpen
  \bibfield  {author} {\bibinfo {author} {\bibfnamefont {M.}~\bibnamefont
  {Trif}}, \bibinfo {author} {\bibfnamefont {V.~N.}\ \bibnamefont {Golovach}},
  \ and\ \bibinfo {author} {\bibfnamefont {D.}~\bibnamefont {Loss}},\
  }\href@noop {} {\bibfield  {journal} {\bibinfo  {journal} {Phys. Rev. B}\
  }\textbf {\bibinfo {volume} {77}},\ \bibinfo {pages} {045434} (\bibinfo
  {year} {2008})}\BibitemShut {NoStop}%
\bibitem [{\citenamefont {Hu}\ \emph {et~al.}(2012)\citenamefont {Hu},
  \citenamefont {Liu},\ and\ \citenamefont {Nori}}]{Hu_PRB_2012}%
  \BibitemOpen
  \bibfield  {author} {\bibinfo {author} {\bibfnamefont {X.}~\bibnamefont
  {Hu}}, \bibinfo {author} {\bibfnamefont {Y.-x.}\ \bibnamefont {Liu}}, \ and\
  \bibinfo {author} {\bibfnamefont {F.}~\bibnamefont {Nori}},\ }\href@noop {}
  {\bibfield  {journal} {\bibinfo  {journal} {Phys. Rev. B}\ }\textbf {\bibinfo
  {volume} {86}},\ \bibinfo {pages} {035314} (\bibinfo {year}
  {2012})}\BibitemShut {NoStop}%
\bibitem [{\citenamefont {Beaudoin}\ \emph {et~al.}(2016)\citenamefont
  {Beaudoin}, \citenamefont {Lachance-Quirion}, \citenamefont {Coish},\ and\
  \citenamefont {Pioro-Ladri\`{e}re}}]{Beaudoin_Nanotech_2017}%
  \BibitemOpen
  \bibfield  {author} {\bibinfo {author} {\bibfnamefont {F.}~\bibnamefont
  {Beaudoin}}, \bibinfo {author} {\bibfnamefont {D.}~\bibnamefont
  {Lachance-Quirion}}, \bibinfo {author} {\bibfnamefont {W.~A.}\ \bibnamefont
  {Coish}}, \ and\ \bibinfo {author} {\bibfnamefont {M.}~\bibnamefont
  {Pioro-Ladri\`{e}re}},\ }\href@noop {} {\bibfield  {journal} {\bibinfo
  {journal} {Nanotechnology}\ }\textbf {\bibinfo {volume} {27}},\ \bibinfo
  {pages} {464003} (\bibinfo {year} {2016})}\BibitemShut {NoStop}%
\bibitem [{\citenamefont {Frey}\ \emph {et~al.}(2012)\citenamefont {Frey},
  \citenamefont {Leek}, \citenamefont {Beck}, \citenamefont {Blais},
  \citenamefont {Ihn}, \citenamefont {Ensslin},\ and\ \citenamefont
  {Wallraff}}]{Walraff_2012_PRL}%
  \BibitemOpen
  \bibfield  {author} {\bibinfo {author} {\bibfnamefont {T.}~\bibnamefont
  {Frey}}, \bibinfo {author} {\bibfnamefont {P.~J.}\ \bibnamefont {Leek}},
  \bibinfo {author} {\bibfnamefont {M.}~\bibnamefont {Beck}}, \bibinfo {author}
  {\bibfnamefont {A.}~\bibnamefont {Blais}}, \bibinfo {author} {\bibfnamefont
  {T.}~\bibnamefont {Ihn}}, \bibinfo {author} {\bibfnamefont {K.}~\bibnamefont
  {Ensslin}}, \ and\ \bibinfo {author} {\bibfnamefont {A.}~\bibnamefont
  {Wallraff}},\ }\href@noop {} {\bibfield  {journal} {\bibinfo  {journal}
  {Phys. Rev. Lett.}\ }\textbf {\bibinfo {volume} {108}},\ \bibinfo {pages}
  {046807} (\bibinfo {year} {2012})}\BibitemShut {NoStop}%
\bibitem [{\citenamefont {Petersson}\ \emph {et~al.}(2012)\citenamefont
  {Petersson}, \citenamefont {McFaul}, \citenamefont {Schroer}, \citenamefont
  {Jung}, \citenamefont {Taylor}, \citenamefont {Houck},\ and\ \citenamefont
  {Petta}}]{Petersson_Nature_2012}%
  \BibitemOpen
  \bibfield  {author} {\bibinfo {author} {\bibfnamefont {K.~D.}\ \bibnamefont
  {Petersson}}, \bibinfo {author} {\bibfnamefont {L.~W.}\ \bibnamefont
  {McFaul}}, \bibinfo {author} {\bibfnamefont {M.~D.}\ \bibnamefont {Schroer}},
  \bibinfo {author} {\bibfnamefont {M.}~\bibnamefont {Jung}}, \bibinfo {author}
  {\bibfnamefont {J.~M.}\ \bibnamefont {Taylor}}, \bibinfo {author}
  {\bibfnamefont {A.~A.}\ \bibnamefont {Houck}}, \ and\ \bibinfo {author}
  {\bibfnamefont {J.~R.}\ \bibnamefont {Petta}},\ }\href@noop {} {\bibfield
  {journal} {\bibinfo  {journal} {Nature}\ }\textbf {\bibinfo {volume} {490}},\
  \bibinfo {pages} {380} (\bibinfo {year} {2012})}\BibitemShut {NoStop}%
\bibitem [{\citenamefont {Viennot}\ \emph {et~al.}(2015)\citenamefont
  {Viennot}, \citenamefont {Dartiailh}, \citenamefont {Cottet},\ and\
  \citenamefont {Kontos}}]{Viennot408}%
  \BibitemOpen
  \bibfield  {author} {\bibinfo {author} {\bibfnamefont {J.~J.}\ \bibnamefont
  {Viennot}}, \bibinfo {author} {\bibfnamefont {M.~C.}\ \bibnamefont
  {Dartiailh}}, \bibinfo {author} {\bibfnamefont {A.}~\bibnamefont {Cottet}}, \
  and\ \bibinfo {author} {\bibfnamefont {T.}~\bibnamefont {Kontos}},\
  }\href@noop {} {\bibfield  {journal} {\bibinfo  {journal} {Science}\ }\textbf
  {\bibinfo {volume} {349}},\ \bibinfo {pages} {408} (\bibinfo {year}
  {2015})}\BibitemShut {NoStop}%
\bibitem [{\citenamefont {Kawakami}\ \emph {et~al.}(2014)\citenamefont
  {Kawakami}, \citenamefont {Scarlino}, \citenamefont {Ward}, \citenamefont
  {Braakman}, \citenamefont {Savage}, \citenamefont {Lagally}, \citenamefont
  {Friesen}, \citenamefont {Coppersmith}, \citenamefont {Eriksson},\ and\
  \citenamefont {Vandersypen}}]{Kawakami.2014}%
  \BibitemOpen
  \bibfield  {author} {\bibinfo {author} {\bibfnamefont {E.}~\bibnamefont
  {Kawakami}}, \bibinfo {author} {\bibfnamefont {P.}~\bibnamefont {Scarlino}},
  \bibinfo {author} {\bibfnamefont {D.~R.}\ \bibnamefont {Ward}}, \bibinfo
  {author} {\bibfnamefont {F.~R.}\ \bibnamefont {Braakman}}, \bibinfo {author}
  {\bibfnamefont {D.~E.}\ \bibnamefont {Savage}}, \bibinfo {author}
  {\bibfnamefont {M.~G.}\ \bibnamefont {Lagally}}, \bibinfo {author}
  {\bibfnamefont {M.}~\bibnamefont {Friesen}}, \bibinfo {author} {\bibfnamefont
  {S.~N.}\ \bibnamefont {Coppersmith}}, \bibinfo {author} {\bibfnamefont
  {M.~A.}\ \bibnamefont {Eriksson}}, \ and\ \bibinfo {author} {\bibfnamefont
  {L.~M.~K.}\ \bibnamefont {Vandersypen}},\ }\href@noop {} {\bibfield
  {journal} {\bibinfo  {journal} {Nat. Nanotechnol.}\ }\textbf {\bibinfo
  {volume} {9}},\ \bibinfo {pages} {666} (\bibinfo {year} {2014})}\BibitemShut
  {NoStop}%
\bibitem [{\citenamefont {Fowler}\ \emph {et~al.}(2012)\citenamefont {Fowler},
  \citenamefont {Mariantoni}, \citenamefont {Martinis},\ and\ \citenamefont
  {Cleland}}]{Fowler_PRA_2012}%
  \BibitemOpen
  \bibfield  {author} {\bibinfo {author} {\bibfnamefont {A.~G.}\ \bibnamefont
  {Fowler}}, \bibinfo {author} {\bibfnamefont {M.}~\bibnamefont {Mariantoni}},
  \bibinfo {author} {\bibfnamefont {J.~M.}\ \bibnamefont {Martinis}}, \ and\
  \bibinfo {author} {\bibfnamefont {A.~N.}\ \bibnamefont {Cleland}},\
  }\href@noop {} {\bibfield  {journal} {\bibinfo  {journal} {Phys. Rev. A}\
  }\textbf {\bibinfo {volume} {86}},\ \bibinfo {pages} {032324} (\bibinfo
  {year} {2012})}\BibitemShut {NoStop}%
\bibitem [{\citenamefont {C\'orcoles}\ \emph {et~al.}(2015)\citenamefont
  {C\'orcoles}, \citenamefont {Magesan}, \citenamefont {Srinivasan},
  \citenamefont {Cross}, \citenamefont {Steffen}, \citenamefont {Gambetta},\
  and\ \citenamefont {Chow}}]{IBM_NatComm_2015}%
  \BibitemOpen
  \bibfield  {author} {\bibinfo {author} {\bibfnamefont {A.~D.}\ \bibnamefont
  {C\'orcoles}}, \bibinfo {author} {\bibfnamefont {E.}~\bibnamefont {Magesan}},
  \bibinfo {author} {\bibfnamefont {S.~J.}\ \bibnamefont {Srinivasan}},
  \bibinfo {author} {\bibfnamefont {A.~W.}\ \bibnamefont {Cross}}, \bibinfo
  {author} {\bibfnamefont {M.}~\bibnamefont {Steffen}}, \bibinfo {author}
  {\bibfnamefont {J.~M.}\ \bibnamefont {Gambetta}}, \ and\ \bibinfo {author}
  {\bibfnamefont {J.~M.}\ \bibnamefont {Chow}},\ }\href@noop {} {\bibfield
  {journal} {\bibinfo  {journal} {Nat. Commun.}\ }\textbf {\bibinfo {volume}
  {6}},\ \bibinfo {pages} {6979} (\bibinfo {year} {2015})}\BibitemShut
  {NoStop}%
\bibitem [{\citenamefont {Debnath}\ \emph {et~al.}(2016)\citenamefont
  {Debnath}, \citenamefont {Linke}, \citenamefont {Figgatt}, \citenamefont
  {Landsman}, \citenamefont {Wright},\ and\ \citenamefont
  {Monroe}}]{Monroe_Nature_2016}%
  \BibitemOpen
  \bibfield  {author} {\bibinfo {author} {\bibfnamefont {S.}~\bibnamefont
  {Debnath}}, \bibinfo {author} {\bibfnamefont {N.~M.}\ \bibnamefont {Linke}},
  \bibinfo {author} {\bibfnamefont {C.}~\bibnamefont {Figgatt}}, \bibinfo
  {author} {\bibfnamefont {K.~A.}\ \bibnamefont {Landsman}}, \bibinfo {author}
  {\bibfnamefont {K.}~\bibnamefont {Wright}}, \ and\ \bibinfo {author}
  {\bibfnamefont {C.}~\bibnamefont {Monroe}},\ }\href@noop {} {\bibfield
  {journal} {\bibinfo  {journal} {Nature}\ }\textbf {\bibinfo {volume} {536}},\
  \bibinfo {pages} {63} (\bibinfo {year} {2016})}\BibitemShut {NoStop}%
\bibitem [{\citenamefont {Majer}\ \emph {et~al.}(2007)\citenamefont {Majer},
  \citenamefont {Chow}, \citenamefont {Gambetta}, \citenamefont {Koch},
  \citenamefont {Johnson}, \citenamefont {Schreier}, \citenamefont {Frunzio},
  \citenamefont {Schuster}, \citenamefont {Houck}, \citenamefont {Wallraff},
  \citenamefont {Blais}, \citenamefont {Devoret}, \citenamefont {Girvin},\ and\
  \citenamefont {Schoelkopf}}]{Majer2007}%
  \BibitemOpen
  \bibfield  {author} {\bibinfo {author} {\bibfnamefont {J.}~\bibnamefont
  {Majer}}, \bibinfo {author} {\bibfnamefont {J.~M.}\ \bibnamefont {Chow}},
  \bibinfo {author} {\bibfnamefont {J.~M.}\ \bibnamefont {Gambetta}}, \bibinfo
  {author} {\bibfnamefont {J.}~\bibnamefont {Koch}}, \bibinfo {author}
  {\bibfnamefont {B.~R.}\ \bibnamefont {Johnson}}, \bibinfo {author}
  {\bibfnamefont {J.~A.}\ \bibnamefont {Schreier}}, \bibinfo {author}
  {\bibfnamefont {L.}~\bibnamefont {Frunzio}}, \bibinfo {author} {\bibfnamefont
  {D.~I.}\ \bibnamefont {Schuster}}, \bibinfo {author} {\bibfnamefont {A.~A.}\
  \bibnamefont {Houck}}, \bibinfo {author} {\bibfnamefont {A.}~\bibnamefont
  {Wallraff}}, \bibinfo {author} {\bibfnamefont {A.}~\bibnamefont {Blais}},
  \bibinfo {author} {\bibfnamefont {M.~H.}\ \bibnamefont {Devoret}}, \bibinfo
  {author} {\bibfnamefont {S.~M.}\ \bibnamefont {Girvin}}, \ and\ \bibinfo
  {author} {\bibfnamefont {R.~J.}\ \bibnamefont {Schoelkopf}},\ }\href@noop {}
  {\bibfield  {journal} {\bibinfo  {journal} {Nature}\ }\textbf {\bibinfo
  {volume} {449}},\ \bibinfo {pages} {443} (\bibinfo {year}
  {2007})}\BibitemShut {NoStop}%
\bibitem [{\citenamefont {Sillanpaa}\ \emph {et~al.}(2007)\citenamefont
  {Sillanpaa}, \citenamefont {Park},\ and\ \citenamefont
  {Simmonds}}]{Sillanpaa2007}%
  \BibitemOpen
  \bibfield  {author} {\bibinfo {author} {\bibfnamefont {M.~A.}\ \bibnamefont
  {Sillanpaa}}, \bibinfo {author} {\bibfnamefont {J.~I.}\ \bibnamefont {Park}},
  \ and\ \bibinfo {author} {\bibfnamefont {R.~W.}\ \bibnamefont {Simmonds}},\
  }\href@noop {} {\bibfield  {journal} {\bibinfo  {journal} {Nature}\ }\textbf
  {\bibinfo {volume} {449}},\ \bibinfo {pages} {438} (\bibinfo {year}
  {2007})}\BibitemShut {NoStop}%
\bibitem [{\citenamefont {Mi}\ \emph {et~al.}(2017{\natexlab{b}})\citenamefont
  {Mi}, \citenamefont {Cady}, \citenamefont {Zajac}, \citenamefont {Stehlik},
  \citenamefont {Edge},\ and\ \citenamefont {Petta}}]{Mi_APL_2016}%
  \BibitemOpen
  \bibfield  {author} {\bibinfo {author} {\bibfnamefont {X.}~\bibnamefont
  {Mi}}, \bibinfo {author} {\bibfnamefont {J.~V.}\ \bibnamefont {Cady}},
  \bibinfo {author} {\bibfnamefont {D.~M.}\ \bibnamefont {Zajac}}, \bibinfo
  {author} {\bibfnamefont {J.}~\bibnamefont {Stehlik}}, \bibinfo {author}
  {\bibfnamefont {L.~F.}\ \bibnamefont {Edge}}, \ and\ \bibinfo {author}
  {\bibfnamefont {J.~R.}\ \bibnamefont {Petta}},\ }\href@noop {} {\bibfield
  {journal} {\bibinfo  {journal} {Appl. Phys. Lett.}\ }\textbf {\bibinfo
  {volume} {110}},\ \bibinfo {pages} {043502} (\bibinfo {year}
  {2017}{\natexlab{b}})}\BibitemShut {NoStop}%
\bibitem [{\citenamefont {Burkard}\ and\ \citenamefont
  {Imamoglu}(2006)}]{Burkard_PRB_2006}%
  \BibitemOpen
  \bibfield  {author} {\bibinfo {author} {\bibfnamefont {G.}~\bibnamefont
  {Burkard}}\ and\ \bibinfo {author} {\bibfnamefont {A.}~\bibnamefont
  {Imamoglu}},\ }\href@noop {} {\bibfield  {journal} {\bibinfo  {journal}
  {Phys. Rev. B}\ }\textbf {\bibinfo {volume} {74}},\ \bibinfo {pages} {041307}
  (\bibinfo {year} {2006})}\BibitemShut {NoStop}%
\bibitem [{\citenamefont {Jin}\ \emph {et~al.}(2012)\citenamefont {Jin},
  \citenamefont {Marthaler}, \citenamefont {Shnirman},\ and\ \citenamefont
  {Sch\"on}}]{Jin_PRL_2012}%
  \BibitemOpen
  \bibfield  {author} {\bibinfo {author} {\bibfnamefont {P.-Q.}\ \bibnamefont
  {Jin}}, \bibinfo {author} {\bibfnamefont {M.}~\bibnamefont {Marthaler}},
  \bibinfo {author} {\bibfnamefont {A.}~\bibnamefont {Shnirman}}, \ and\
  \bibinfo {author} {\bibfnamefont {G.}~\bibnamefont {Sch\"on}},\ }\href@noop
  {} {\bibfield  {journal} {\bibinfo  {journal} {Phys. Rev. Lett.}\ }\textbf
  {\bibinfo {volume} {108}},\ \bibinfo {pages} {190506} (\bibinfo {year}
  {2012})}\BibitemShut {NoStop}%
\bibitem [{\citenamefont {Taylor}\ \emph {et~al.}(2013)\citenamefont {Taylor},
  \citenamefont {Srinivasa},\ and\ \citenamefont {Medford}}]{Taylor_PRL_2013}%
  \BibitemOpen
  \bibfield  {author} {\bibinfo {author} {\bibfnamefont {J.~M.}\ \bibnamefont
  {Taylor}}, \bibinfo {author} {\bibfnamefont {V.}~\bibnamefont {Srinivasa}}, \
  and\ \bibinfo {author} {\bibfnamefont {J.}~\bibnamefont {Medford}},\
  }\href@noop {} {\bibfield  {journal} {\bibinfo  {journal} {Phys. Rev. Lett.}\
  }\textbf {\bibinfo {volume} {111}},\ \bibinfo {pages} {050502} (\bibinfo
  {year} {2013})}\BibitemShut {NoStop}%
\bibitem [{\citenamefont {Russ}\ and\ \citenamefont
  {Burkard}(2015)}]{Guido_RX_PRB_2015}%
  \BibitemOpen
  \bibfield  {author} {\bibinfo {author} {\bibfnamefont {M.}~\bibnamefont
  {Russ}}\ and\ \bibinfo {author} {\bibfnamefont {G.}~\bibnamefont {Burkard}},\
  }\href@noop {} {\bibfield  {journal} {\bibinfo  {journal} {Phys. Rev. B}\
  }\textbf {\bibinfo {volume} {92}},\ \bibinfo {pages} {205412} (\bibinfo
  {year} {2015})}\BibitemShut {NoStop}%
\bibitem [{\citenamefont {{Probst}}\ \emph {et~al.}(2017)\citenamefont
  {{Probst}}, \citenamefont {{Bienfait}}, \citenamefont {{Campagne-Ibarcq}},
  \citenamefont {{Pla}}, \citenamefont {{Albanese}}, \citenamefont {{Da Silva
  Barbosa}}, \citenamefont {{Schenkel}}, \citenamefont {{Vion}}, \citenamefont
  {{Esteve}}, \citenamefont {{M{\o}lmer}}, \citenamefont {{Morton}},
  \citenamefont {{Heeres}},\ and\ \citenamefont
  {{Bertet}}}]{Bertet_Arxiv_2017}%
  \BibitemOpen
  \bibfield  {author} {\bibinfo {author} {\bibfnamefont {S.}~\bibnamefont
  {{Probst}}}, \bibinfo {author} {\bibfnamefont {A.}~\bibnamefont
  {{Bienfait}}}, \bibinfo {author} {\bibfnamefont {P.}~\bibnamefont
  {{Campagne-Ibarcq}}}, \bibinfo {author} {\bibfnamefont {J.~J.}\ \bibnamefont
  {{Pla}}}, \bibinfo {author} {\bibfnamefont {B.}~\bibnamefont {{Albanese}}},
  \bibinfo {author} {\bibfnamefont {J.~F.}\ \bibnamefont {{Da Silva Barbosa}}},
  \bibinfo {author} {\bibfnamefont {T.}~\bibnamefont {{Schenkel}}}, \bibinfo
  {author} {\bibfnamefont {D.}~\bibnamefont {{Vion}}}, \bibinfo {author}
  {\bibfnamefont {D.}~\bibnamefont {{Esteve}}}, \bibinfo {author}
  {\bibfnamefont {K.}~\bibnamefont {{M{\o}lmer}}}, \bibinfo {author}
  {\bibfnamefont {J.~J.~L.}\ \bibnamefont {{Morton}}}, \bibinfo {author}
  {\bibfnamefont {R.}~\bibnamefont {{Heeres}}}, \ and\ \bibinfo {author}
  {\bibfnamefont {P.}~\bibnamefont {{Bertet}}},\ }\href@noop {} {\bibfield
  {journal} {\bibinfo  {journal} {arXiv:1708.09287}\ } (\bibinfo {year}
  {2017})}\BibitemShut {NoStop}%
\bibitem [{\citenamefont {Nadj-Perge}\ \emph {et~al.}(2010)\citenamefont
  {Nadj-Perge}, \citenamefont {Frolov}, \citenamefont {Bakkers},\ and\
  \citenamefont {Kouwenhoven}}]{Nadj_Nature_2010}%
  \BibitemOpen
  \bibfield  {author} {\bibinfo {author} {\bibfnamefont {S.}~\bibnamefont
  {Nadj-Perge}}, \bibinfo {author} {\bibfnamefont {S.~M.}\ \bibnamefont
  {Frolov}}, \bibinfo {author} {\bibfnamefont {E.~P. A.~M.}\ \bibnamefont
  {Bakkers}}, \ and\ \bibinfo {author} {\bibfnamefont {L.~P.}\ \bibnamefont
  {Kouwenhoven}},\ }\href@noop {} {\bibfield  {journal} {\bibinfo  {journal}
  {Nature}\ }\textbf {\bibinfo {volume} {468}},\ \bibinfo {pages} {1084}
  (\bibinfo {year} {2010})}\BibitemShut {NoStop}%
\bibitem [{\citenamefont {Zajac}\ \emph {et~al.}(2015)\citenamefont {Zajac},
  \citenamefont {Hazard}, \citenamefont {Mi}, \citenamefont {Wang},\ and\
  \citenamefont {Petta}}]{Dave_DQD_APL}%
  \BibitemOpen
  \bibfield  {author} {\bibinfo {author} {\bibfnamefont {D.~M.}\ \bibnamefont
  {Zajac}}, \bibinfo {author} {\bibfnamefont {T.~M.}\ \bibnamefont {Hazard}},
  \bibinfo {author} {\bibfnamefont {X.}~\bibnamefont {Mi}}, \bibinfo {author}
  {\bibfnamefont {K.}~\bibnamefont {Wang}}, \ and\ \bibinfo {author}
  {\bibfnamefont {J.~R.}\ \bibnamefont {Petta}},\ }\href@noop {} {\bibfield
  {journal} {\bibinfo  {journal} {Appl. Phys. Lett.}\ }\textbf {\bibinfo
  {volume} {106}},\ \bibinfo {eid} {223507} (\bibinfo {year}
  {2015})}\BibitemShut {NoStop}%
\bibitem [{\citenamefont {Schuster}\ \emph {et~al.}(2005)\citenamefont
  {Schuster}, \citenamefont {Wallraff}, \citenamefont {Blais}, \citenamefont
  {Frunzio}, \citenamefont {Huang}, \citenamefont {Majer}, \citenamefont
  {Girvin},\ and\ \citenamefont {Schoelkopf}}]{Schuster_ACStark_2005}%
  \BibitemOpen
  \bibfield  {author} {\bibinfo {author} {\bibfnamefont {D.~I.}\ \bibnamefont
  {Schuster}}, \bibinfo {author} {\bibfnamefont {A.}~\bibnamefont {Wallraff}},
  \bibinfo {author} {\bibfnamefont {A.}~\bibnamefont {Blais}}, \bibinfo
  {author} {\bibfnamefont {L.}~\bibnamefont {Frunzio}}, \bibinfo {author}
  {\bibfnamefont {R.-S.}\ \bibnamefont {Huang}}, \bibinfo {author}
  {\bibfnamefont {J.}~\bibnamefont {Majer}}, \bibinfo {author} {\bibfnamefont
  {S.~M.}\ \bibnamefont {Girvin}}, \ and\ \bibinfo {author} {\bibfnamefont
  {R.~J.}\ \bibnamefont {Schoelkopf}},\ }\href@noop {} {\bibfield  {journal}
  {\bibinfo  {journal} {Phys. Rev. Lett.}\ }\textbf {\bibinfo {volume} {94}},\
  \bibinfo {pages} {123602} (\bibinfo {year} {2005})}\BibitemShut {NoStop}%
\bibitem [{\citenamefont {Elzerman}\ \emph {et~al.}(2004)\citenamefont
  {Elzerman}, \citenamefont {Hanson}, \citenamefont {Willems~van Beveren},
  \citenamefont {Witkamp}, \citenamefont {Vandersypen},\ and\ \citenamefont
  {Kouwenhoven}}]{Elzerman_Readout}%
  \BibitemOpen
  \bibfield  {author} {\bibinfo {author} {\bibfnamefont {J.~M.}\ \bibnamefont
  {Elzerman}}, \bibinfo {author} {\bibfnamefont {R.}~\bibnamefont {Hanson}},
  \bibinfo {author} {\bibfnamefont {L.~H.}\ \bibnamefont {Willems~van
  Beveren}}, \bibinfo {author} {\bibfnamefont {B.}~\bibnamefont {Witkamp}},
  \bibinfo {author} {\bibfnamefont {L.~M.~K.}\ \bibnamefont {Vandersypen}}, \
  and\ \bibinfo {author} {\bibfnamefont {L.~P.}\ \bibnamefont {Kouwenhoven}},\
  }\href@noop {} {\bibfield  {journal} {\bibinfo  {journal} {Nature}\ }\textbf
  {\bibinfo {volume} {430}},\ \bibinfo {pages} {431} (\bibinfo {year}
  {2004})}\BibitemShut {NoStop}%
\end{thebibliography}%

\end{document}